\documentclass{aa}  

\usepackage{graphicx}
\usepackage{txfonts}
\usepackage{makecell}
\usepackage[colorlinks=true, linkcolor=blue, citecolor=blue, urlcolor=blue]{hyperref}
\usepackage{xcolor}
\begin{document}

   \title{Predicting galaxy bias using machine learning}

   \author{Catalina Riveros-Jara\thanks{cata.riverosj@gmail.com}\inst{1}
          \and
          Antonio D. Montero-Dorta\thanks{amonterodorta@gmail.com}\inst{1}
          \and 
          Natália V. N. Rodrigues\inst{2}
          \and
          Pía Amigo\inst{1}
          \and
          Natalí S. M. de Santi\inst{3,4}
          \and
          Andrés Balaguera-Antolínez
          \and 
          Raul Abramo\inst{5}
          \and
          Neill Guzmán\inst{1}
          \and
          M. Celeste Artale\inst{6}
          }

   \institute{Departamento de Física, Universidad Técnica Federico Santa María,
              Avenida Vicuña Mackenna 3939, San Joaquín, Santiago, Chile
         \and
         HEP Division, Argonne National Laboratory, 9700 South Cass Avenue, Lemont, IL 60439, USA
         \and
         Berkeley Center for Cosmological Physics, University of California, Berkeley,
         341 Campbell Hall, Berkeley, CA 94720, U.S.A.
         \and
         Physics Division, Lawrence Berkeley National Laboratory, 1 Cyclotron Road, Berkeley, CA 94720, U.S.A.
        \and 
        Departamento de Física Matemática, Instituto de Física, Universidade de São Paulo, Rua do Matão 1371, CEP 05508-090 São Paulo, Brazil
         \and
        Universidad Andres Bello, Facultad de Ciencias Exactas, Departamento de Fisica y Astronomia, Instituto de Astrofisica, Fernandez Concha 700, Las Condes, Santiago RM, Chile
        }

   \date{Received September 15, 1996; accepted March 16, 1997}

 
  \abstract
    {Understanding how galaxies trace the underlying matter density field is essential for characterizing the influence of the large-scale structure on galaxy formation, being therefore a key ingredient in observational cosmology. This connection, commonly described through the galaxy bias, $b$, can be studied effectively using machine-learning (ML) techniques, which offer strong predictive capabilities and can capture non-linear relationships in high-dimensional data. Recent work has also highlighted the need for probabilistic methods to properly account for the intrinsic stochasticity of  this connection.}
   {We aim to incorporate the linear bias parameter assigned to individual galaxies into a ML framework, quantify its dependence on various halo and environmental properties, and evaluate whether different algorithms can accurately predict this parameter and reproduce the scatter in several bias relations.} 
   {We use data from the IllustrisTNG300 magnetohydrodynamical simulation, including the distance to different cosmic web structures computed with DisPerSE. These data are complemented with an object-by-object estimator of the large-scale linear bias ($b_i$), providing the individual contribution of each galaxy to the bias of the entire population. Our ML framework uses three models to predict $b_i$: a Random Forest Regressor, a single-output Neural Network and a probabilistic method (Normalizing Flows).}
   {We recover the full hierarchy of galaxy bias dependencies, showing that the most informative features are the overdensities, particularly $\delta_8$, followed by the distances to cosmic-web structures and selected internal halo properties, most notably the formation redshift ($z_{1/2}$). We also demonstrate that Normalizing Flows clearly outperform deterministic methods in predicting galaxy bias, including its joint distributions with galaxy properties, owing to their ability to capture the intrinsic variance associated with the stochastic nature of the matter-halo-galaxy connection. 
   Our ML framework provides a foundation for future efforts to measure individual bias with upcoming spectroscopic surveys.}
   {}

   \keywords{galaxies: statistics --
                dark matter --
                large-scale structure of Universe -- 
                galaxy: halo--
                galaxy: evolution --
                galaxy: formation
               }

   \maketitle
%
\newcommand{\aba}[1]{\textcolor{teal}{(ABA: #1)}}
\newcommand{\natali}[1]{\textcolor{purple}{[natali: #1]}}
\newcommand{\natalia}[1]{\textcolor{cyan}{[natalia: #1]}}

\section{Introduction}
From a cosmological perspective, it is often useful to view the Universe as composed of three interconnected layers: the continuous matter density field, the discrete field of dark matter (DM) halos, and the discrete field of galaxies, which form through the cooling and condensation of gas within the halos' potential wells \citep{White1978}. Characterizing the mapping between these cosmic fields has key implications for models of galaxy formation and for the extraction of cosmological information from galaxy surveys, motivating the ongoing development of innovative analytical, computational, and theoretical techniques \citep{Wechsler2018}.

In this context, machine learning (ML) has emerged as a powerful approach for capturing non-linear relationships in high-dimensional data, improving the efficiency and accuracy of simulations, emulators, and cosmological inference as a whole (e.g., \citealt{LucieSmith2018, Peel2019, VillaescusaNavarro2021,Kreisch2022,Perez2023}). ML has also become an integral part of the standard toolkit for exploring the halo-galaxy connection, not only for its ability to reproduce the complex relationships between galaxies and DM halos, but also for its potential to shed light on the physical mechanisms underlying this fundamental link \citep{Jo2019, Stiskalek2022, de_Santi_2022, Jespersen2022, Rodrigues_2023, Lovell2023, Sullivan_2023,   Chuang2024, Rodrigues2025}. Importantly, recent studies have emphasized the need for a probabilistic framework to predict galaxy properties based on halo and environmental properties, reflecting the inherent stochasticity of the underlying physical processes \citep{Rodrigues_2023, 2023A&A...673A.130B,Rodrigues2025}.

The bias of halos and galaxies --relating their overdensities and clustering to those of the underlying matter density field-- provides an analytical bridge between the aforementioned cosmic fields \citep[e.g.][]{Kaiser1984, Efstathiou1988, Mo1996, Desjacques2018}. In this framework, the linear halo bias is often defined as the ratio of power spectra (or correlation functions), $b = P_{\mathrm{hm}}(k)/P_{\mathrm{mm}}(k)$\footnote{In this expression, $P_{\mathrm{hm}}(k)$ and $P_{\mathrm{mm}}(k)$ represent the halo-matter and matter-matter power spectra, respectively.}, which on large scales recovers the expectation from peak--background split theory for a given population of objects (see e.g. \citealt{Paranjape2012, Schmidt2013}).  It has also been recently shown that it is possible to isolate the individual contribution of each object to the global bias of the population \citep{Paranjape2018, Han2019, Stucker2025}. In this work, we leverage the analytical advantages of an object-by-object formalism to incorporate bias into a ML-based reproduction framework to investigate the dependence of galaxy bias on halo and environmental properties.

In this context, halo bias is known to depend primarily on the peak height of density fluctuations, $\nu$ -- and consequently, the halo mass -- as can be analytically derived from structure formation formalisms (see, e.g., \citealt{Kaiser1984,Bardeen1986,Mo1996, ShethTormen1999, Sheth2001, Tinker2010}). At fixed $\nu$ or halo mass, however, halo bias has been shown to depend on a variety of additional internal halo properties, including formation time, concentration, spin, or shape (see, e.g., \citealt{Sheth2004,gao2005,Wechsler2006,Gao2007,Dalal2008, Angulo2008,Li2008,faltenbacher2010, Lazeyras2017,2018Salcedo,Mao2018, Han2018,SatoPolito2019, Johnson2019, Paranjape2018, Ramakrishnan2019,Contreras2019, Tucci2021,  Contreras2021_cosmo, MonteroDorta2021, MonteroRodriguez2024, Balaguera2024A, MonteroDorta2025}). As it can be also inferred from analytical models of structure formation, halo bias is also naturally connected to the environment. At fixed halo mass, the dependencies on the local density across multiple scales, on the geometry of the tidal field and its anisotropic (traceless) component, and on the specific location of haloes within the cosmic web have been extensively characterized in simulations (e.g., \citealt{Borzyszkowski2017, Musso2018, Paranjape2018, Ramakrishnan2019, 2023A&A...673A.130B,MonteroRodriguez2024, Balaguera2024A}). 

The intrinsic relation between galaxies and halos translates into galaxy bias, which is naturally dominated by the bias of the galaxies' hosting halos. The dependence of galaxy clustering on galaxy properties and environment has been extensively investigated using spectroscopic surveys, showing that galaxies of higher mass, redder colors, early-type morphologies and those in denser regions are more tightly clustered across a wide range of scales than the rest of the population (e.g., \citealt{AbbasSheth2006,Meneux2006,Coil2006, Abbas2007, Skibba2009, Zehavi2011, Hartley2013, McNaughtRoberts2014, Zhai2023}). These general trends mostly persist when the analysis is performed at fixed stellar mass (e.g., \citealt{Li2006, Zehavi2011, LawSmith2017}), and even group mass \citep{Wang2008, Rodriguez2026}.

In order to evaluate the predictive power of ML in terms of reproducing galaxy bias and uncovering its dependencies on halo and environmental properties, we employ several models applied to the \textsc{IllustrisTNG}\footnote{\url{http://www.tng-project.org}.} hydrodynamical simulation, following a methodology similar to that of \citet{de_Santi_2022, Rodrigues_2023, Rodrigues2025}. We test both deterministic models, such as random forests regressor (RF) and neural networks (NN), and probabilistic approaches, such as normalizing flows (NFs), to demonstrate the ability of the latter to naturally capture the stochasticity of the linear bias\footnote{For simplicity, by stochasticity in this context we mean the intrinsic variance in the sample, once the different properties employed to reproduce the bias are taken into account.}. Developing this framework is particularly valuable for measuring galaxy bias, a key quantity at the interface between astrophysics and cosmology that not only encodes the relationship between galaxies and dark matter, but also carries cosmological information.

The paper is organized as follows. Section~\ref{sec:Data} describes the IllustrisTNG data with the distances computed with DisPerSE used in this work, in addition to the method for computing the galaxy-by-galaxy individual bias. The ML models employed in this study, along with the selected evaluation metrics are specified in Sec.~\ref{sec:Machine Learning Framework}. In Sec.~\ref{sec:Correlation Between Features} we analyze the dependence of $b_i$ on all selected features. Section~\ref{sec:Predicting Galaxy Bias} exhibits the predictions of the three ML algorithms with their performance and the individual probability distributions obtained with NFs. The main results of this work are summarized in Sec.~\ref{sec:Discussion and Conclusions}, together with an interpretation of the key findings. 

\section{Data}
\label{sec:Data}
\subsection{IllustrisTNG}
Our study employs data from the IllustrisTNG suite of magnetohydrodynamical cosmological simulations (hereafter referred to as TNG; \citealt{Pillepich2018b,Pillepich2018,Nelson2018_ColorBim,Nelson2019,Marinacci2018,Naiman2018,Springel2018}). These simulations were carried out with the {\sc arepo} moving-mesh code \citep{Springel2010}, which numerically solves the equations of magnetohydrodynamics coupled with self-gravity, and is considered a major improvement over the original Illustris framework \citep{Vogelsberger2014a, Vogelsberger2014b, Genel2014}. The updated TNG sub-grid physics models include processes such as star formation, radiative cooling with metal-line contributions, chemical enrichment from Type II and Type Ia supernovae as well as AGB stars, and feedback from both stellar populations and supermassive black holes. These physical prescriptions were calibrated to reproduce key observables, including the $z=0$ stellar mass function, the cosmic star formation rate density, halo gas fractions, galaxy size distributions, and the black hole–stellar mass relation (see aforementioned references for more detailed information).

For the purposes of this work, which focuses on studying a large-scale property, we adopt the TNG300-1 simulation (referred to as TNG300\footnote{\url{https://www.tng-project.org/data/docs/specifications/}.} throughout this paper), as it provides a great trade-off between resolution and cosmological volume within the TNG suite, making it a suitable choice for this scenario. TNG300 is the largest box among the TNG simulations—a periodic cube with a side length of $205,h^{-1}$ Mpc. It follows the evolution of $2500^3$ DM particles (each with mass $4.0\times10^7 h^{-1} {\rm M_\odot}$) and an equal number of initial gas cells (each of mass $7.6\times10^6 h^{-1} {\rm M_\odot}$). The IllustrisTNG300 simulation adopts the standard $\Lambda$ cold dark matter ($\Lambda$CDM) cosmology \citep{Planck2016}, with parameters $\Omega_{\rm m} = 0.3089$, $\Omega_{\rm b} = 0.0486$, $\Omega_\Lambda = 0.6911$, $H_0 = 100 \, h\, {\rm km \, s^{-1}Mpc^{-1}}$ with $h=0.6774$, $\sigma_8 = 0.8159$, and $n_s = 0.9667$. This simulation has proven to be a powerful tool for investigating galaxy formation and the connection between DM halos and galaxies, and it has contributed to a wide range of scientific studies (e.g., \citealt{Springel2018,Pillepich2018,Beltz-Mohrmann2020,Contreras2020,Gu2020,Hadzhiyska2020,Hadzhiyska2021,Shi2020,MonteroDorta2020B,MonteroDorta2020C,MonteroDorta2021,Engler2021,MonteroDorta2023}).

DM halos in TNG are identified using a friends-of-friends (FOF) algorithm with a linking length equal to 0.2 times the mean inter-particle separation \citep{Davis1985}. Gravitationally bound substructures (termed subhalos) are detected using the {\sc subfind} algorithm \citep{Springel2001,Dolag2009}. Subhalos containing a non-zero stellar component are classified as galaxies.

In this work, we restrict the analysis to central galaxies in the TNG300 catalog, which simplifies the modeling of galaxy clustering in terms of the selected halo and environmental properties. To exclude nonphysical values within the dataset, we apply two selection cuts: $\log_{10}\text{M}_{\text{vir}}[h^{-1}\text{M}_\odot] > 10.5$ and $\log_{10}\text{M}_{*}[h^{-1}\text{M}_\odot] > 8.75$, for halo and stellar mass, respectively. These thresholds ensure that each halo contains more than 500 DM particles and each galaxy at least 50 stellar particles. After these selections, the final sample consists of 174,607 objects.

\subsubsection{Internal halo and galaxy properties}
Several halo and galaxy properties from TNG300 are considered in this work. For halos, we selected the following features:

\begin{itemize}
    \item Virial mass ($M_{\text{vir}}\,[h^{-1}\text{M}_\odot]$), defined as the total mass enclosed within a sphere whose mean density equals 200 times the critical density, and computed in TNG by adding up the mass of all the gas cells contained within a sphere of radius $R_{\text{vir}}$.
    \item Age ($z_{\,\text{form}}$), defined as the redshift at which half of the present-day halo mass has been accreted into a single subhalo for the first time, reason why this age is described in terms of the formation redshift ($z_{\text{1/2}}$). To compute this, as in \cite{de_Santi_2022}, the progenitors of the main branch of the subhalo merger tree determined using {\sc sublink} (which is initialized at $z=6$) were considered.
    \item Concentration ($c_{\text{vir}}$), defined as the ratio between the virial radius ($R_{\text{vir}}$) and the scale radius ($R_\text{s}$):
    \begin{equation}
        c_{\text{vir}} = \frac{R_{\text{vir}}}{R_{\text{s}}},
    \end{equation}
    where $R_{\text{s}}$ is derived by fitting a Navarro-Frenk-White dark matter density profile \citep{Navarro_1997} to individual halos.
    \item Spin ($\lambda_{\text{halo}}$), defined as in \cite{Bullock_2001}, namely:
    \begin{equation}
        \centering
        \lambda_{\text{halo}} = \frac{|J|}{\sqrt{2}M_{\text{vir}}V_{\text{vir}}R_{\text{vir}}},
    \end{equation}
    where $J$ and $V_{\text{vir}}$ correspond to the angular momentum of the halo and its circular velocity at $R_{\text{vir}}$ respectively.
\end{itemize}

For the galaxies (i.e., subhalos with non-zero stellar components in TNG), we considered the following properties:
\begin{itemize}
    \item Stellar mass ($M_{\ast}\,[h^{-1}\text{M}_\odot]$), defined as the total mass of all stellar particles gravitationally bound to each subhalo.
    \item Galaxy color (g - i), computed using the magnitudes provided by IllustrisTNG. These are obtained by summing the luminosities of all stellar particles from each subhalo (see \citealp{Buser1978}). The TNG magnitudes are intrinsic, i.e., they do not include attenuation due to dust.
\end{itemize}

\subsubsection{Local environment and cosmic web}

In this work, we employ multiple environmental properties to characterize both the local environment and the location of galaxies within the structures that make up the cosmic web.
Regarding the local environment, we use the overdensities around halos
on scales of 3, 5 and 8 $h^{\text{-1}}\text{Mpc}$, i.e.,  $\delta_{\text{3}},\, \delta_{\text{5}}$ and $\delta_{\text{8}}$, respectively. These overdensities are defined as the number of subhalos within a sphere, center at each halo, of radius equal to the corresponding scale, normalized by the total number density of subhalos in the TNG300 box (e.g., \citealt{Artale2018,Bose2019}). 

The position of galaxies within the cosmic web is characterized by means of their distance to the critical points of the density field provided by the Discrete Persistent
Structures Extractor (DisPerSE; \citealp{Sousbie2011a}) TNG300 public catalog \citep{Duckworth2020a,Duckworth2020b}\footnote{The catalog can be directly downloaded from the TNG webpage. It can also be accessed from the GitHub link \url{https://github.com/Chris-Duckworth/disperse_TNG}.}. DisPerSE is a computational framework designed to automatically identify persistent topological structures such as peaks, voids, walls, and particularly filaments from the density field using Morse theory. The catalog employs a value of $\sigma=4$ for the ``persistence'' parameter without any additional smoothing; for more information, we refer the reader to the aforementioned works (see also, e.g., \citealt{MonteroRodriguez2024}).

Each galaxy in our parent catalog was assigned a distance to the closest critical point corresponding to each type, namely:

\begin{itemize}
    \item $d_{\text{min}}\:  \text{ckpc}/h$: Distance to the nearest minimum  (roughly corresponding to voids centers).

    \item $d_{\text{node}}\:  \text{ckpc}/h$: Distance to the nearest maximum (nodes).
    
    \item $d_{\text{saddle 1}}\:  \text{ckpc}/h$ and $d_{\text{saddle 2}}\:  \text{ckpc}/h$: Distance to the nearest n-saddle point ($\text{n} = 1,2$). Saddle 1 corresponds to a critical point where one dimension is collapsing (wall saddles), while saddle 2 corresponds to a critical point where two dimensions are collapsing (filament saddles).
    
    \item $d_{\text{skel}}\:  \text{ckpc}/h$: Distance to the nearest filament segment.
\end{itemize}

\subsubsection{Large scale bias}
\label{sec:large-scale bias}
The large-scale linear bias characterizes, in a statistical sense, the relation between the spatial distribution of galaxies (or DM halos) and the underlying matter density field. It is commonly defined as the ratio between correlation functions or power spectra. For instance, in configuration space one can write $b = \xi_{\text{gm}}/\xi_{\text{mm}}$, where $\xi_{\text{mm}}$ denotes the auto-correlation of the DM density field, and $\xi_{\text{gm}}$ represents the cross-correlation between galaxies and DM. Analogously, the bias can also be inferred from ratios of auto- or cross-power spectra (see, e.g., \citealt{pollack2012}). A limitation of these traditional approaches is that bias measurements are typically obtained for halo subsamples selected according to a given property of interest. This subdivision can significantly reduce the signal-to-noise ratio of the measured power spectra, making them increasingly dominated by shot noise. Moreover, it restricts the range of Fourier modes available for the analysis, an effect that is particularly severe for rare tracers such as high-mass halos. As a result, disentangling the specific dependence of bias on halo internal or environmental properties becomes challenging. 

To address this issue, a convenient object-by-object large-scale bias estimator was introduced by \citet{Paranjape2018}. This method exploits basic properties of the discrete Fourier transform to build an estimator whose ensemble average reproduces the trends measured by conventional large-scale estimators (\citealt{pollack2012}). Within this framework, the effective large-scale bias of a galaxy sample is interpreted as the mean of a distribution of individual bias values, each corresponding to a single galaxy or halo. This or similar individual-bias approaches have been successfully employed in several works (e.g., \citealt{Han2018, Ramakrishnan2019, Contreras2021a, Balaguera2024A, Balaguera2024B, MonteroDorta2025, MonteroDorta2025_voids}), which have shown that the mean individual halo bias as a function of halo mass (the so-called \emph{bias function}) agrees remarkably well with both traditional measurements \citep[e.g.][]{Tinker2010} and theoretical predictions based on the peak-height formalism of density fluctuations (e.g., \citealt{Mo1996}). 

In this study, we adopt the philosophy of \citet{Paranjape2018} to estimate the large-scale linear bias for each galaxy in our TNG300 sample. This approach allows us to assign a bias value to each galaxy in the sample and include it as one of its individual properties, among the other extracted features. Unlike standard techniques that deliver population-averaged values, this method provides a conceptually straightforward and flexible way to analyze galaxy clustering as a function of various halo and environmental features.

Following the formalism of \citet{Balaguera2024A}, the individual linear bias of a galaxy $i$ located at position $\mathbf{r}$, $b_i$, is computed as
\begin{equation}
b_i =\frac{\sum_{j,k_{j}<k_{max}}N^{j}_{k}\langle \exp[-i\bf{k}\cdot \bf{r}_{i}]\delta_{\mathrm{DM}}^{*}(\bf{k}) \rangle_{k_{j}}}{\sum_{j,k_{j}<k_{max}} N^{j}_{k}P_{\rm DM}(k_{j})},
   \label{eq:bias}
\end{equation}
where $\delta_{\mathrm{DM}}^{*}(\mathbf{k})$ is the Fourier transform of the DM density contrast field, $P_{\rm DM}(k_{j})$ is the matter power spectrum, and $N^{j}_{k}$ is the number of Fourier modes in the $j$-th spherical shell\footnote{The assignment of individual galaxy bias has been performed using the \texttt{CosmiCCcodes} library at \url{https://github.com/balaguera/CosmicCodes}.}. A comprehensive derivation and interpretation of Eq.~\eqref{eq:bias} can be found in \citet{Paranjape2018}.

The sum in Eq.~\eqref{eq:bias} extends over the range of wavenumbers where the ratio between galaxy and dark matter power spectra remains constant. Given the simulation volume, we adopt 
$k_{\rm max} \leq 0.2 \,h \, \rm{Mpc}^{-1}$, up to which this ratio remains consistent with a constant value. Within this framework, the effective large-scale bias of a population containing $N_{\rm G}$ galaxies is simply obtained, as mentioned above, as the mean of their individual biases:
\begin{equation}
    \langle b \rangle_{\rm G} = \frac{\sum_{i = 1, N_{\rm G}} b_i}{N_{\rm G}},
   \label{eq:bias_mean}
\end{equation}

It is important to emphasize that Eq.~\eqref{eq:bias} does not describe a local galaxy–matter connection. Instead, it quantifies the contribution of each galaxy to the large-scale bias of the sample (Eq.~\ref{eq:bias_mean}).


\section{Machine learning framework}
\label{sec:Machine Learning Framework}
In this work, we aim to predict the individual linear bias of galaxies in TNG300 based on internal halo properties and local-environment features along with cosmic web distances\footnote{See \cite{Balaguera2024B} for an inverse approach to the one described here, in which bias is used to assign halo properties in a probabilistic manner without relying on ML.}. For this purpose, the performance of three different ML models is carefully evaluated. 

\subsection{Models}
\label{sec:models}
To cover a representative range of options, we compare the performance of two different single-point deterministic estimators to that of a probabilistic approach. In the case of the deterministic models, we employ a RF and a fully connected NN with a single output. As a prominent example of a probabilistic model, we use NFs.   

RF \citep{Breiman2001} is a supervised learning algorithm, built as an ensemble of regression trees to obtain a more robust and accurate model. RFs introduce randomness through two mechanisms: \emph{bootstrap sampling} and \emph{random feature selection}. This randomness increases robustness and generalization, as well as reduces overfitting. Each tree is trained on a bootstrap sample drawn from the training set, and at each node the algorithm randomly selects a subset of features (without replacement) to determine the best split, forcing the trees to use different combinations of features. The predictions from all trees are then aggregated to produce the final result. Owing to its simplicity and interpretability, this model has been increasingly applied in extragalactic astrophysics and cosmology (e.g., \citealt{carliles2007photometricredshiftestimationsdss}; \citealt{Wang2013}; \citealt{Qiu_2024}; \citealt{Bluck_2025}). In this work, we employ the \texttt{RandomForestRegressor} implementation from the \texttt{scikit-learn} library \citep{pedregosa2018scikitlearnmachinelearningpython}.

NNs are a Deep Learning technique widely applied to a variety of tasks, including both regression and classification. They consist of interconnected nodes (neurons) organized into layers, where each node has an associated weight ($\omega$) and bias ($b$), which are used to process the input and pass the result to the next layer through a nonlinear transformation called the activation function (e.g., ReLU, LeakyReLU, SiLU). Both $\omega$ and $b$ are iteratively adjusted to minimize a loss function, chosen here to be the mean squared error (MSE), which quantifies the difference between the true targets ($y^{\text{true}}$) and the predicted values ($y^{\text{pred}}$). To update the parameters (weights and biases), we employ the Adam optimizer \citep{kingma2017adammethodstochasticoptimization}. Our models are implemented using the \texttt{PyTorch} library \citep{paszke2019pytorchimperativestylehighperformance}. Recent applications of NNs in this context include \citet{Calderon_2019}, \citet{de_Santi_2022}, \citet{Shao_2022}, and \citet{Rodrigues_2023}.

Following \cite{Rodrigues2025}, we adopt a NF as a probabilistic, generative framework to model galaxy clustering. NFs learn a bijective transformation that maps samples from a simple base distribution $p_Z(z)$, typically a standard Gaussian, to a complex target distribution $p_X(x)$ through a sequence of invertible and differentiable mappings, with probability densities related via the change-of-variables formula: 
\begin{equation}
    p_X(x) = p_Z(z) \Bigg| \text{det} \left( \frac{\partial f(z)}{\partial z} \right)\Bigg|^{-1} =  p_Z(f^{-1}(x)) \Bigg|\text{det} \left(\frac{\partial f^{-1}(x)}{\partial x}\right)\Bigg| ~.
    \label{eq:change of variables form}
\end{equation}
In this work we choose $p_Z(z)$ to be a standard Normal distribution. These transformations are commonly parameterized using affine functions \citep{papamakarios2018maf} or monotonic spline bijections \citep{durkan2019, dolatabadi2020}, whose parameters are learned by NNs.

To efficiently handle high-dimensional distributions, NFs employ structured transformations such as coupling and autoregressive flows. Coupling layers partition the variables into subsets, where one subset parametrizes the transformation of the other, while autoregressive flows (\citealt{kingma2017}) generalize this approach by modeling each dimension sequentially, yielding a fully factorized conditional structure. This framework naturally extends to conditional density estimation by conditioning the transformation parameters on additional inputs.

In this work, we implement neural spline flows to model the conditional distribution $p(b_i|\{x\}_i)$, where $\{x\}_i$ denotes the set of halo and environmental properties associated with that particular galaxy.
The model is optimized via the negative log-likelihood loss function:
\begin{equation}
    \mathcal{L} = - \frac{1}{N}\sum_i^N \log p(b_i|\{x\}_i) ~,
    \label{eq:negative log-like}
\end{equation}
where $N$ is the number of instances in the dataset. The inference is then performed by sampling the Gaussian base distribution and pushing these samples forward through the conditional transformation, parameterized by the input features ($\{x\}$), to give the posterior samples.

For higher-dimensional conditioned outputs, we employ neural spline autoregressive flows to jointly predict multiple galaxy properties conditioned on the set of halo and environment properties. The model is implemented using the \texttt{Pyro} library \citep{bingham2018}.

\subsection{Evaluation metrics}
\label{sec:evaluation metrics}

When assessing the performance of a ML model, it is essential to quantify how close the predicted values ($y^{\text{pred}}$) are to the true ones ($y^{\text{true}}$). This can be achieved using a set of evaluation metrics. In this work, we evaluate the performance of each model on the test set using three metrics: the Pearson Correlation Coefficient (PCC), the Kolmogorov–Smirnov test (K–S test), and the Wasserstein Distance (WD). These metrics provide distinct but complementary perspectives: the PCC estimates linear correlation, the K–S test evaluates distributional similarity, and the WD measures global similarity between the entire predicted and true distributions.

The PCC, defined in Eq.~\eqref{eq:pcc} below, provides a normalized measure of covariance, yielding values between $-1$ and $1$. It quantifies the degree of linear correlation between the predicted and true values, where $+1$ indicates a perfect positive linear relation and $-1$ a perfect negative one:
\begin{equation}
\text{PCC} = \frac{\text{cov}\left(y^{\text{pred}}, y^{\text{true}} \right)}{\sigma_{y^{\text{pred}}} \sigma_{y^{\text{true}}}}.
\label{eq:pcc}
\end{equation}
A higher PCC value corresponds to a stronger agreement between predictions and targets in terms of linear trends.

For the 1-D K–S test, we adopted the statistic $D$
computed using the \texttt{scipy.stats.ks\_2samp} function from the \texttt{scipy.stats} module \citep{Virtanen_2020}. This statistic quantifies the maximum difference between the cumulative distribution functions (CDFs) of two distributions, $f_1(x_1)$ and $f_2(x_2)$, namely:
\begin{equation}
D = \text{max} \left(|F_1(x_1) - F_2(x_2)| \right),
\end{equation}
where $F_1(x_1) = \text{CDF}[f_1(x_1)]$ and $F_2(x_2) = \text{CDF}[f_2(x_2)]$. This test evaluates whether two samples are drawn from the same underlying distribution; larger $D$ values indicate greater discrepancies between them. For the 2-D K-S Test, we used the \citep{Taillon_2DKS} repository.

Finally, the WD (also known as the Earth Mover’s Distance) is a measure of the similarity between two probability distributions. It can be interpreted as the minimum “cost” required to transform one distribution into the other through optimal transport. Given two probability distributions $u$ and $v$, defined over $\mathcal{R}$, the one-dimensional Wasserstein distance is expressed as:
    \begin{equation}
        W(u,v) = \inf_{\pi \in \Gamma(u,v)} \int_{\mathcal{R} \times \mathcal{R}} |x-y| \, d\pi(x,y),
        \label{eq: W distance}
    \end{equation}
where $\Gamma(u,v)$ represents the set of all joint probability distributions $\pi$ on $\mathcal{R} \times \mathcal{R}$ with marginals $u (x)$ and $v(x)$ (\citealt{ramdas2015wassersteinsampletestingrelated, Virtanen_2020}). Intuitively, this formulation seeks the joint distribution $\pi(x,y)$ that minimizes the expected distance between points $x$ and $y$ when the values $(x,y)$ are sampled from $\pi(x,y)$. The WD accounts for the global structure of the distributions, making it particularly sensitive to differences in both location and shape. Consequently, lower WD values indicate a better match between the predicted and true distributions.
This metric is widely employed in optimal transport theory and generative modeling applications (e.g., \citealp{arjovsky2017wassersteingan}). In this work, we compute the 1-D Wasserstein Distance using the \texttt{SciPy} library \citep{Virtanen_2020}, and the 2-D WD using \cite{Flamary_2021}.

These metrics (PCC, K-S Test and WD) are evaluated using point estimates (i.e., $y^{\text{pred}}$ is directly compared to $y^{\text{true}}$). Since NFs predict the full posterior, it is necessary to summarize the posterior into a single number. Then, for the NF model, we computed the metrics in two ways. First, $y^{\text{pred}}$ was taken as a single random realization selected from the sampled outputs. Second, we used the mean value on all realizations, denoted as "avg". 

To evaluate the accuracy and calibration of the probabilistic predictions produced by the NF model, we performed a Test for Accuracy of Ranked Posteriors (TARP\footnote{This test only applies for generative methods.}; \citealt{lemos2023}), which quantifies whether the posterior distributions predicted by the model are statistically well calibrated, i.e. whether the true values are contained within the predicted credibility intervals at the expected rates. This test can be found in Appendix \ref{appendix:Apendix_2}.

\section{Correlation between features}
\label{sec:Correlation Between Features}
It is expected that many of the properties that we analyze in this work display a certain degree of correlation between them, and also with respect to linear bias. 
We assess the degree of correlation among features using the PCC, which quantifies the linear relationship between pairs of properties. This constitutes an important step, as ML algorithms do not rely on prior physical assumptions. As a means of providing a general view of the data set, we evaluate here how each property relates to $b_i$.

\begin{figure}
    \centering
    \includegraphics[width=0.8\linewidth]{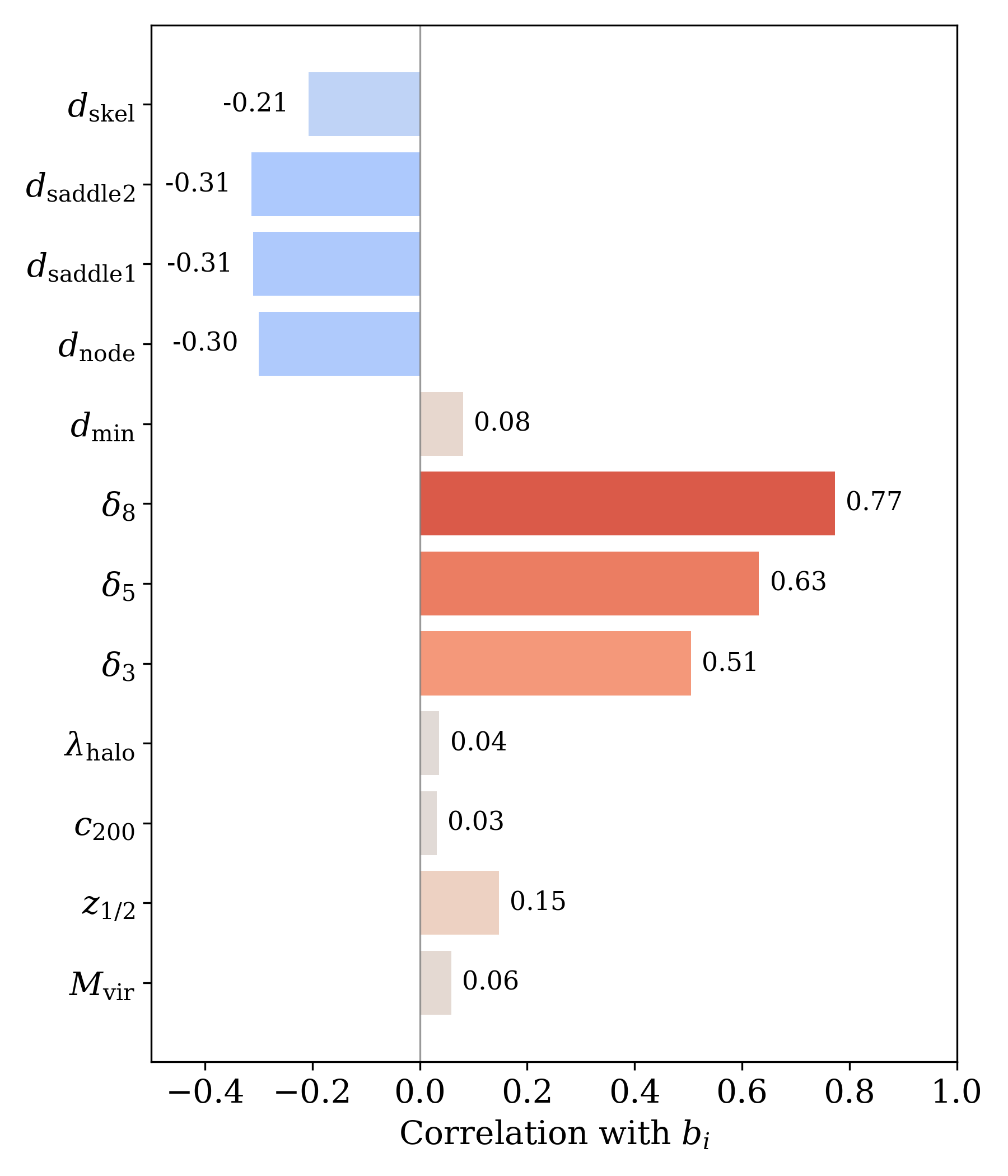}
    \caption{Correlation of each selected property with individual galaxy bias, considering the entire mass range. Each bar is accompanied with the value of the linear correlation (PCC) between both features.}
    \label{fig:corr_matrix_resumen}
\end{figure}

Figure~\ref{fig:corr_matrix_resumen} shows the correlation of each selected property with the individual galaxy bias across the entire halo mass range. As expected, the local environment exhibits a greater connection with galaxy clustering compared to the remaining properties, with $\delta_8$ being the property most strongly correlated with the bias. For the cosmic web distances, an anti-correlation is observed for all distances except for that to the nearest void. This behavior is not surprising, as we approach denser environments (e.g., filaments and nodes), the bias increases, whereas proximity to less dense regions leads to weaker clustering \citep{MonteroDorta2024,Wang2024, Rodriguez2016}.

Regarding the internal properties of halos, due to intrinsic scatter presented in the bias parameter, the correlation between this parameter and the internal properties of halos is also expected to be small\footnote{\cite{MonteroDorta2020B} showed that, albeit small, these correlations are enough to produce significant level of galaxy assembly bias, understood there as the secondary dependencies of galaxy bias at fixed halo mass.} \citep{Paranjape2018, Han2018, MonteroDorta2020B, Balaguera2024A}, which is consistent with the values presented in Fig.~\ref{fig:corr_matrix_resumen}. Interestingly, the internal property that presents the stronger correlation with galaxy bias is $z_{1/2}$, instead of halo mass. This can be explained by the fact that the measurement, when taken globally for the entire population, is dominated by the more abundant low-mass halos, for which the dependence on halo mass is known to be weak (e.g., \citealt{Tinker2010}). Note also that there is an intrinsic connection between halo mass and age, which can be understood in terms of halo assembly history, where more massive halos—associated with stronger galaxy clustering—tend to form later through hierarchical merging.

\begin{figure}
    \centering
    \includegraphics[width=0.98\linewidth]{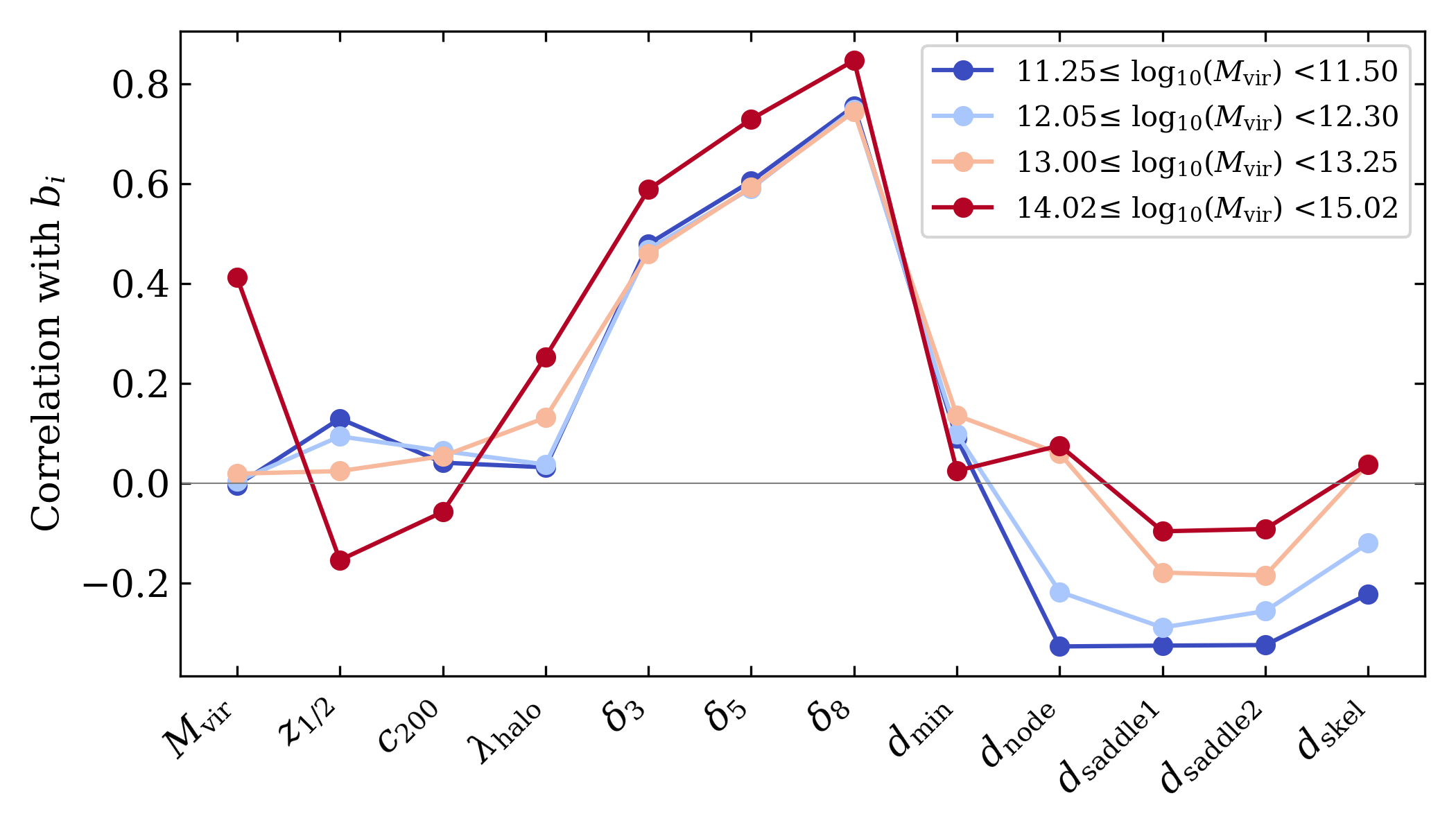}
    \caption{Correlation coefficient between $b_i$ and each selected feature within four different halo mass bins. The bins span 0.25 dex, except for the most massive bin.}
    \label{fig:corr_matrix_bins_puntos}
\end{figure}

The relation between halo mass and large-scale linear bias has been widely studied, both analytically and in simulations \citep{Kaiser1984,Bardeen1986,Mo1996, ShethTormen1999, Sheth2001, Tinker2010}. To explore additional dependencies, it is convenient to fix the halo mass. A practical approach to achieve this is by dividing the sample into narrow mass bins, as shown in Fig.~\ref{fig:corr_matrix_bins_puntos}, where the following bins are considered:
$11.25 \leq \log_{10} M_{\text{vir}}[h^{-1}\text{M}_\odot]<11.50$;
$12.05 \leq \log_{10} M_{\text{vir}}[h^{-1}\text{M}_\odot]<12.30$;
$13.00 \leq \log_{10} M_{\text{vir}}[h^{-1}\text{M}_\odot]<13.25$;
$14.02 \leq \log_{10} M_{\text{vir}}[h^{-1}\text{M}_\odot]<15.02$. All these bins span 0.25 dex, except for the most massive one, which is broader due to the limited number of objects in this region within the TNG300 simulation box. This simple exercise illustrates the power of individual bias to capture correlations at fixed halo mass, an effect commonly referred to as secondary halo bias (or halo assembly bias). The complete set of correlations between all features can be seen in the correlation matrices of Fig.~\ref{fig:corr_matrix_bins} in Appendix~\ref{appendix:Apendix_1}.

At fixed halo mass, the correlation between the individual galaxy bias and halo mass increases for the most massive bin, consistent with the bias–mass relation. Furthermore, $\delta_8$ remains the property most strongly correlated with galaxy bias across all mass bins. As halo mass increases, the correlation between galaxy clustering and the formation redshift of the host halos decreases, whereas the opposite trend is observed for the spin, consistent with the well-known halo assembly bias and spin bias trends, respectively (see, e.g., \citealt{SatoPolito2019}). In addition, decreasing halo mass strengthens the correlation between individual galaxy bias and proximity to different cosmic web structures, except for voids, in agreement with \cite{MonteroRodriguez2024}.

The trends observed in both Figures~\ref{fig:corr_matrix_resumen} and \ref{fig:corr_matrix_bins_puntos} are consistent with theoretical expectations and previous studies (see, e.g., \citealt{SatoPolito2019,Han2018, Paranjape2018,alam2019,MonteroDorta2020B, MonteroDorta2024,Balaguera2024A}), and form the foundation of the predictions of this work, as we want to test whether the different ML algorithms employed can successfully capture the underlying physical relationships among the various features. In addition, we also address feature importance using permutational feature importance in Sec. \ref{sec:model_comp}, allowing us to compare the relevance assigned by each model with the trends identified in this correlation analysis.

\section{Predicting galaxy bias}
\label{sec:Predicting Galaxy Bias}
\subsection{Training procedure and model optimization}
\label{sec:Training Procedure and Model Optimization}
From the total sample of 174,607 objects used in this analysis, 58\% were assigned to the training set, 30\% to the test set, and 12\% to the validation set. To determine the optimal architecture of each model, we employed the \texttt{OPTUNA} framework \citep{Akiba_2019} to perform Bayesian optimization using the Tree-structured Parzen Estimator (TPE; \citealt{Bergstra_2011}), with 100 trials per model.

For the RF, the explored hyperparameters included the number of trees in the forest (N. estimators), the maximum depth of the trees (Max. depth), the minimum number of samples required to split an internal node (Min. samples split), and the maximum number of leaf nodes allowed (Max. leaf nodes). The final set of hyperparameters corresponded to the configuration that achieved the lowest MSE on the validation set.

For the NN, we optimized the number of hidden layers, the layer sizes (i.e., number of nodes per layer), the activation function, the learning rate of the Adam optimizer, and the dropout rate. For the activation function, we considered ReLU, SiLU and LeakyReLU during the optimization process, with the latter yielding the best performance. Similarly, for the NF model, we explored the number of hidden layers, layer sizes, number of spline segments (bins), and batch size, and we sampled 1000 realizations from the learned conditional probability distributions. For both NN and NF models, the final hyperparameters were selected as those yielding the lowest value of the corresponding loss function (see Sec.~\ref{sec:models}) in the validation set. The final set of hyperparameters for the three models are presented in Tables \ref{table:hyperparameters RF}, \ref{table:hyperparameters NN}, and \ref{table:hyperparameters NF}.

\begin{table}[h!]
\caption{Explored hyperparameters for the RF, along with the searched range and the values of the best trial.}                 
\label{table:hyperparameters RF}    
\centering                     
\begin{tabular}{c c c}      
\hline\hline               
RF Hyperparameters & Range of Values & Best Trial\\       
\hline                 
   N. Estimators & [1000, 3000] & 1809 \\   
   Max. Depth & [100, 1000] & 237\\
   Min. Samples Split & [10, 100] &10\\
   Max. Leaf Nodes & [1000, 2000] & 1975\\
\hline                              
\end{tabular}
\end{table}

\begin{table}[h!]
\caption{Explored hyperparameters for the NN, along with the searched range and the values of the best trial.}                 
\label{table:hyperparameters NN}    
\centering                     
\begin{tabular}{c c c}      
\hline\hline               
NN Hyperparameters & Range of Values & Best Trial\\       
\hline                 
N. Layers & [2, 10] & 2\\   
Layers Sizes & [32, 256] & 224, 32\\
Dropout Rate & [0.1, 0.5] & 1.554e-1\\
Learning Rate & [1e-4, 1e-1] & 8.703e-3\\
\hline                              
\end{tabular}
\end{table}

\begin{table}[h!]
\caption{Explored hyperparameters for the NFs, along with the searched range and the values of the best trial.}                 
\label{table:hyperparameters NF}    
\centering                     
\begin{tabular}{c c c}      
\hline\hline               
NF Hyperparameters & Range of Values & Best Trial\\       
\hline                 
   N. Layers & [1, 5] & 3\\   
   Layers Sizes & [4, 128] & 21, 51, 122\\
   Learning Rate & [1e-5, 1e-1] & 1.301e-3\\
   Batch Size & [100, 1000] & 650\\
   Bins & [4, 64] & 46\\
\hline                              
\end{tabular}
\end{table}

\subsection{Model comparison}
\label{sec:model_comp}

\begin{figure*}
    \centering
    \includegraphics[width=1\textwidth]{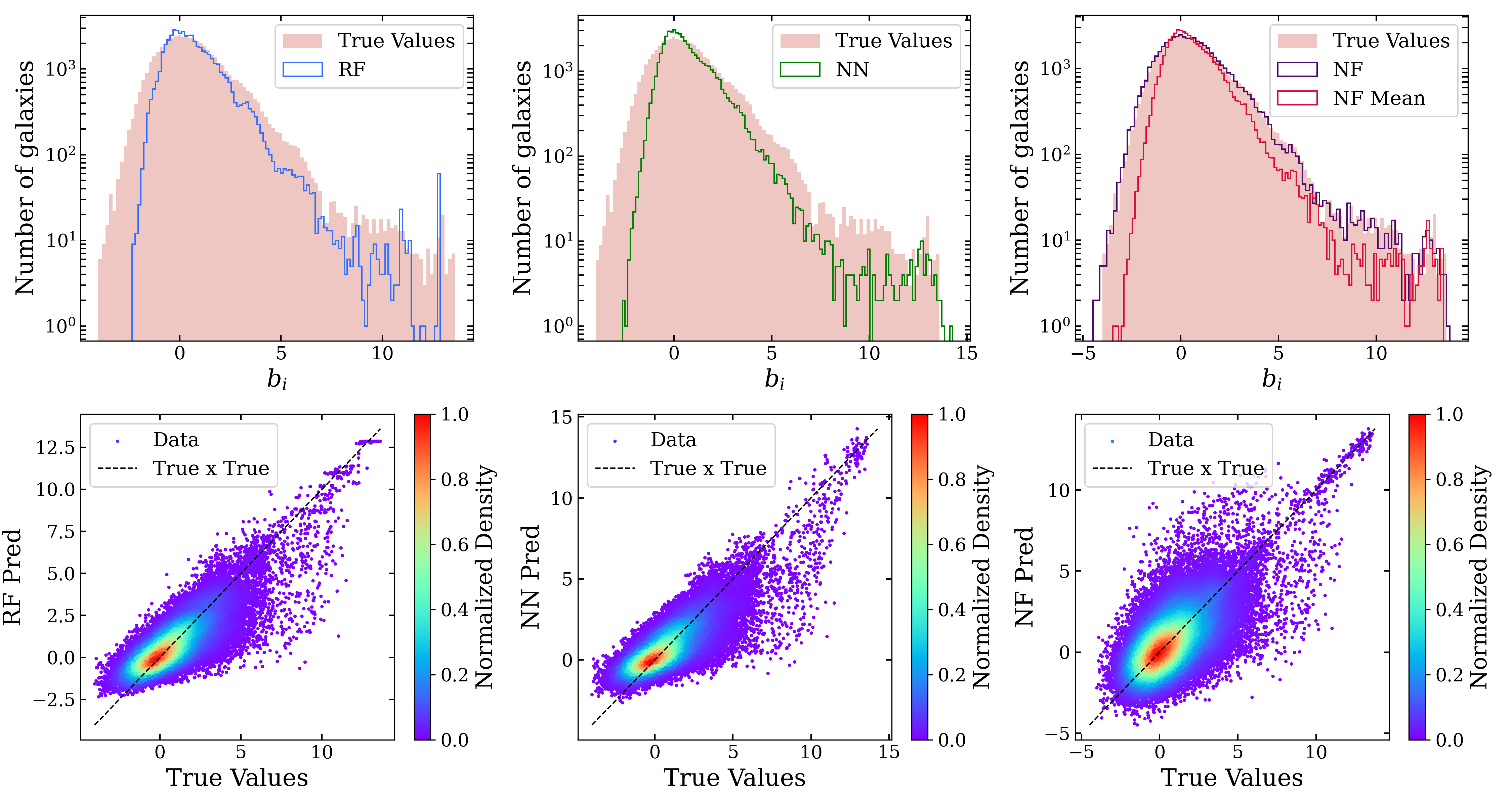}
    \caption{Performance of each model in predicting the individual galaxy bias. The top row shows the true bias values distribution (in pink), along with the predictions from each model in different colors. The NF predictions present both a random realization (NF) and the expected value across the entire sample of realizations (NF Mean). The bottom row displays scatter plots comparing the predicted values (by each model) as a function of the true values, colored by normalized density. Notably, the bottom-right plot emphasizes the use of the NF random realization. The black dashed lines represent the ideal case where predicted values match the true values.}
    \label{fig:3_models_comparison}
\end{figure*}

Figure~\ref{fig:3_models_comparison} presents the predictions from the three models compared to the true values. The top row shows the distributions of the predicted and true individual galaxy bias values for each model, while the bottom row displays scatter plots of predicted versus true values, color-coded by normalized density, which was estimated using Gaussian Kernel Density Estimation (KDE) implemented through the \texttt{scipy.stats.gaussian\_kde} function \citep{Virtanen_2020}.

It is evident that the deterministic models (RF and NN) are unable to closely reproduce the tails of the true bias values histogram, while increasing their precision in the high-density regions where most galaxies are found (see the top row of Fig.~\ref{fig:3_models_comparison}). This behavior is consistent with the corresponding scatter plots (bottom row of Fig.~\ref{fig:3_models_comparison}), where regions of higher normalized density lie closer to the ideal one-to-one relation, while other regions are more dispersed or deviate from the black dashed line.

For the generative model, selecting a single realization\footnote{Each realization is obtained by drawing one sample from the conditional distribution $p(b_i|\{x\}_i)$. Fixing a single realization corresponds to one possible draw from the population-level distribution.} per galaxy reproduces the true bias values distribution more faithfully (see the purple line in the top row of Fig.~\ref{fig:3_models_comparison}). In contrast, taking the mean value across the entire sampled catalog for each galaxy fails to accurately reproduce the histogram of true values, as it approximates a deterministic model (i.e., the resulting distribution, shown in crimson in Fig.~\ref{fig:3_models_comparison}, has a similar shape to those of the NN and RF). This is the same behavior observed in the halo–galaxy connection previously identified and addressed in \citealt{de_Santi_2022} using the SMOGN augmentation technique, and in \cite{Rodrigues_2023, Rodrigues2025} using binning classification schemes with NNs and generative models such as NFs. In Fig.~\ref{fig:3_models_comparison}, both cases (random realization and expected value) are shown alongside the true distribution, while the scatter plot corresponds to the single random realization case (purple line in the histogram).

We also computed the evaluation metrics described in Sec.~\ref{sec:evaluation metrics} for each model on the test set to enable a quantitative comparison. The results for the deterministic estimators are presented in Table~\ref{table:Metrics deterministic}, while those for the probabilistic approach are listed in Table~\ref{table:Metrics probabilistic}. In the latter, "avg" refers to metrics computed using the mean values, whereas the metrics without "avg" correspond to the case in which a single realization is selected.

\begin{table}[h!]
\caption{Metric values for each single-point estimator (RF and NN).}                 
\label{table:Metrics deterministic}    
\centering                     
\begin{tabular}{c c c c}      
\hline\hline               
Model & PCC & K-S Test & WD\\       
\hline                 
   RF & 0.811 & 0.099 & 0.316\\   
   NN & 0.797 & 0.123 & 0.365\\
\hline                              
\end{tabular}
\end{table}

\begin{table}[h!]
\caption{Metric values for the probabilistic approach (NF).}                 
\label{table:Metrics probabilistic}    
\centering                     
\begin{tabular}{c c c c c c }      
\hline\hline               
PCC & K-S Test & WD & PCC$_{\text{avg}}$ & K-S Test$_{\text{avg}}$ & WD$_{\text{avg}}$\\       
\hline                 
   0.652 & 0.010 & 0.046 & 0.807 & 0.086 & 0.318\\   
\hline                              
\end{tabular}
\end{table}

For the deterministic models, RF exhibits slightly better performance than the NN across all three evaluation metrics, indicated by a higher PCC and lower values of the K–S test and WD. In the case of the NF model, when considering a single random realization, both the K–S test and the WD indicate that the predicted and true distributions are remarkably similar. This can be observed by comparing the pink histogram and the purple line in Fig.~\ref{fig:3_models_comparison}, demonstrating the ability of this probabilistic approach to accurately reproduce the true individual bias distribution.

However, selecting a single random realization from the NF-predicted conditional distributions also results in a low PCC value. This is evident in the corresponding scatter plot in Fig.~\ref{fig:3_models_comparison}, where the values are widely spread around the ideal case (black dashed line). Nevertheless, this scatter is highly symmetric, indicating that the model does not exhibit a systematic tendency to overestimate or underestimate the bias values, and that it is able to capture the global behavior of the data. When the expected value over the full set of 1000 sampled realizations is computed, the PCC increases and, when considering all three evaluation metrics, the model achieves a performance comparable to that of the RF.

Altogether, these results suggest that, although both deterministic models are robust in terms or their predictive capability, their predictions remain less accurate, capturing well the mean but not the width of the distributions. Their performance could be likely due to the intrinsic stochasticity of the bias parameter, reinforcing the notion that probabilistic methods are better suited for modeling and reproducing galaxy bias.

We also examined the relative impact of each input feature for the three ML algorithms used to predict galaxy bias. To this end, we employed permutation feature importance (PFI), performing 50 permutations per feature for each model on the test set. This method quantifies the contribution of an individual feature by measuring the degradation in the performance of the model when the relationship between that feature and the target is ``broken". A key advantage of PFI is that it is model-agnostic, allowing it to be applied consistently across different algorithms.

In this section, we consider a feature to be important when its PFI value is high, meaning that shuffling its values leads to a significant decrease in the performance of the model, in other words, the model strongly depends on that feature to make accurate predictions. The full set of importance values for each model is listed in Table~\ref{table:feature importance summary}, where the values are normalized such that the total importance for each model sums to unity.

All three models identify environmental properties as more relevant than the internal properties of the halos for predicting galaxy bias. In particular, $\delta_8$ emerges as the most influential feature, in agreement with the correlation analysis presented in Sec.~\ref{sec:Correlation Between Features}, and exhibits a significant higher importance than the remaining variables. This pronounced difference between the importance of $\delta_8$ and that of the other features further suggests that the models could potentially achieve acceptable predictive performance using only this single property.

Some features exhibit negative importance values, indicating that the performance of the model slightly improves when those variables are shuffled. This behavior suggests that such features do not contribute meaningful information to the predictions and may instead act as noise, and could therefore be discarded in future analyses. It is important to note that, because $\delta_8$ has such a strong impact, even when other features are permuted the algorithms still count on $\delta_8$, which provides most of the relevant information.

Consequently, although all features show some level of correlation with galaxy bias (as discussed in Sec.~\ref{sec:Correlation Between Features}), and between them (as it can be observed in the full correlation matrices shown in Appendix~\ref{appendix:Apendix_2}), shuffling any variable other than $\delta_8$ has only a minor effect on performance. This explains both the low (or negative) PFI values for the remaining features and why the resulting importance ranking does not exactly follow the correlation hierarchy shown in Fig.~\ref{fig:corr_matrix_resumen} (i.e., the properties in Table~\ref{table:feature importance summary} do not follow the same order as in Fig.~\ref{fig:corr_matrix_resumen}, from the most correlated features to those with lower correlation values).

Finally, an interesting aspect of NFs is that this technique enables the prediction of joint probability distributions. While deterministic estimators can also model such distributions, they generally do so with less fidelity than generative methods. To complement our analysis, we predicted pairs of properties and examined the performance of NFs in reproducing the scatter in these relations. As shown in Appendix~\ref{sec:multiproperty predictions}, NFs successfully reproduce 
the scatter in the bias–stellar mass and bias–color relations, yielding a 2D K–S statistic of 0.017 and a 2D WD of 0.030 for the $b_i$–$\log_{10}(M_*[h^{-1}\mathrm{M}_\odot])$ relation, and a 2D K–S statistic of 0.029 and a 2D WD of 0.052 for the $b_i$–$(g-i)$ relation. This scatter at fixed stellar mass and color reflects the intrinsic complexity of the physical processes governing halo and galaxy formation and evolution.

\begin{table}[h!]
\centering
\small
\setlength{\tabcolsep}{3pt}  
\caption{Feature importance values for the three ML models employed in this study: RF, NN, and NF. The features are sorted from highest to lowest importance in each case.}
\label{table:feature importance summary}
\begin{tabular}{c c c c c c}
\hline\hline
\multicolumn{2}{c}{RF} & \multicolumn{2}{c}{NN} & \multicolumn{2}{c}{NF} \\
Feature & Importance & Feature & Importance & Feature & Importance\\
\hline
$\delta_8$ & 0.889 & $\delta_8$ & 0.941 & $\delta_8$ & 0.816 \\
$\delta_5$ & 0.046 & $\delta_5$ & 0.015 & $\delta_5$ & 0.041 \\
$d_{\mathrm{min}}$ & 0.017 & $d_{\mathrm{min}}$ & 0.013 & $d_{\mathrm{node}}$ & 0.038 \\
$d_{\mathrm{node}}$ & 0.015 & $d_{\mathrm{saddle1}}$ & 0.012 & $d_{\mathrm{min}}$ & 0.032 \\
$d_{\mathrm{skel}}$ & 0.012 & $d_{\mathrm{skel}}$ & 0.009 & $d_{\mathrm{saddle1}}$ & 0.024 \\
$d_{\mathrm{saddle1}}$ & 0.011 & $d_{\mathrm{saddle2}}$ & 0.004 & $d_{\mathrm{saddle2}}$ & 0.024 \\
$d_{\mathrm{saddle2}}$ & 0.006 & $d_{\mathrm{node}}$ & 0.004 & $d_{\mathrm{skel}}$ & 0.022 \\
$\delta_3$ & 0.003 & $\delta_3$ & 0.002 & $\delta_3$ & 0.002 \\
$\lambda_{\mathrm{halo}}$ & 1.4$\cdot 10^{-4}$ & $z_{1/2}$ & 1.9$\cdot 10^{-4}$ & $z_{1/2}$ & 6.4$\cdot 10^{-4}$ \\
$c_{200}$ & 1.0$\cdot 10^{-4}$ & $M_{\mathrm{vir}}$ & 1.8$\cdot 10^{-4}$ & $M_{\mathrm{vir}}$ & 5.0$\cdot 10^{-5}$ \\
$M_{\mathrm{vir}}$ & 9.0$\cdot 10^{-5}$ & $c_{200}$ & 1.0$\cdot 10^{-5}$ & $\lambda_{\mathrm{halo}}$ & 4.0$\cdot 10^{-5}$ \\
$z_{1/2}$ & 5.0$\cdot 10^{-5}$ & $\lambda_{\mathrm{halo}}$ & -2.0$\cdot 10^{-5}$ & $c_{200}$ & 2.0$\cdot 10^{-5}$ \\
\hline
\end{tabular}
\end{table}


\begin{figure*}
    \centering
    \includegraphics[width=0.8\textwidth]{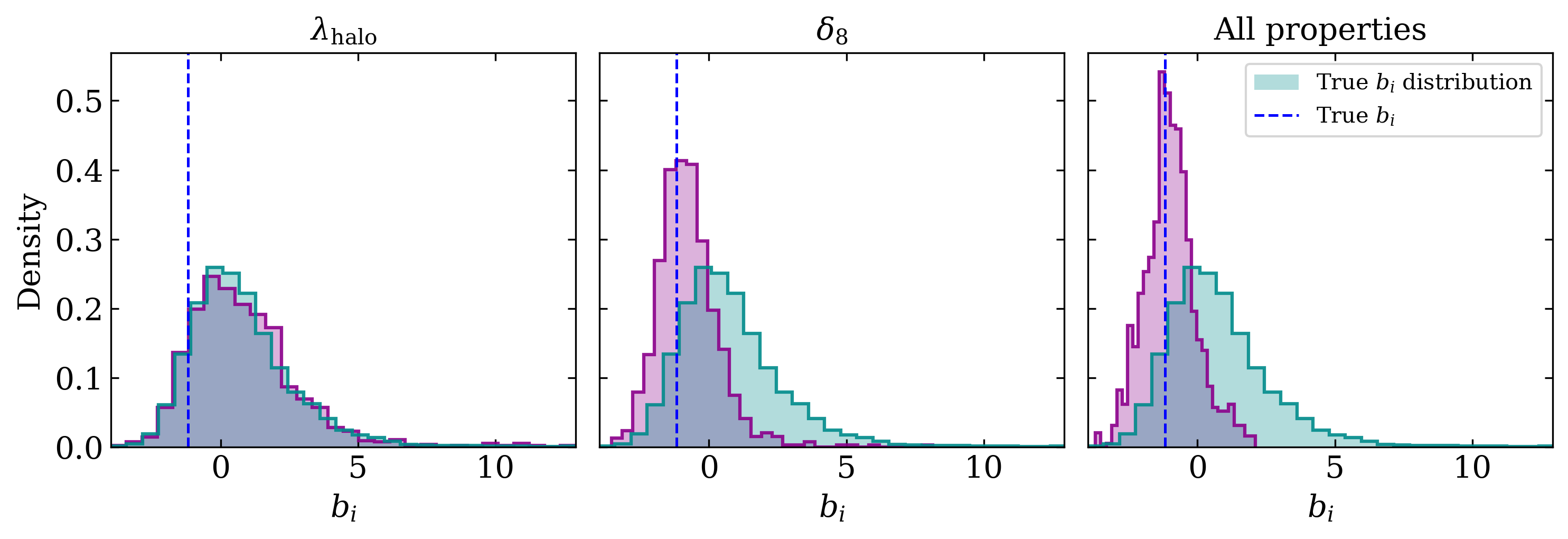}
    \caption{Individual probability distributions (in purple) for one random galaxy in the sample obtained with the NF model trained with three different inputs. The left panel corresponds to the predicted values using $\lambda_{\text{halo}}$ as input, the center panel to the case only using $\delta_8$ as input while the right panel to the case using the complete set of properties described in Sec.~\ref{sec:Data}. The blue dashed line corresponds to the true individual bias value of the galaxy ($b_i = -1.19$), while the cyan distributions show the true bias values as shown in pink in Fig.~\ref{fig:3_models_comparison}.}
    \label{fig:distributions_NF_onegalaxy}
\end{figure*}

\begin{figure}
    \centering
    \includegraphics[width=0.8\linewidth]{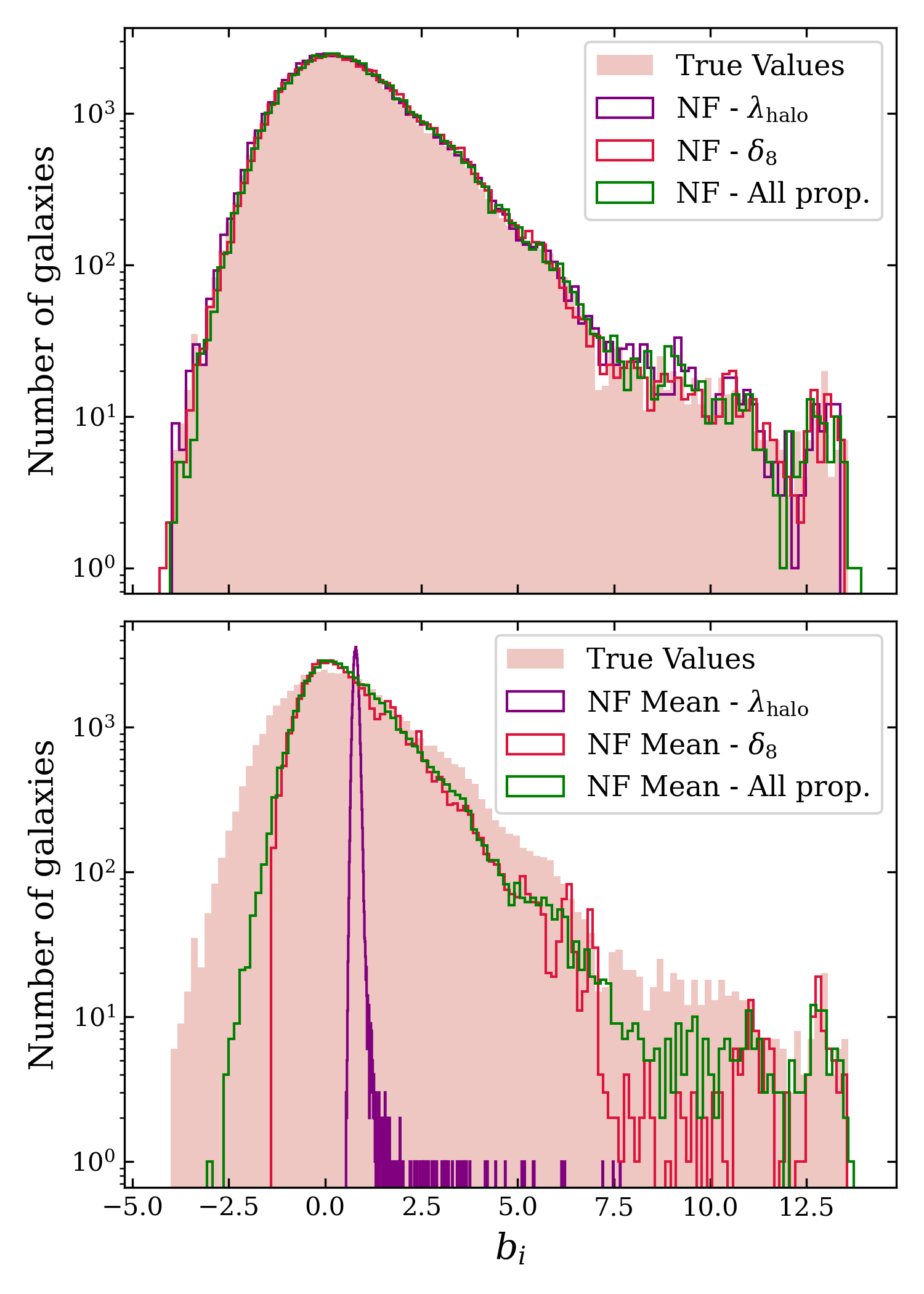}
    \caption{Predicted individual probability distributions obtained with the NF model when using different input properties (only $\lambda_{\text{halo}}$ in purple, only $\delta_8$ in crimson and the full set of halo and environmental properties in green). The top panel shows the distributions obtained when considering one random realization from the entire set of sampled values from the conditional probability distributions, while the bottom panel shows the case when taking the mean values of the individual probability distributions of each galaxy. }
    \label{fig:different_distributions_NF}
\end{figure}

\subsection{Individual probability distributions}
One of the most useful features of NFs is their probabilistic treatment of the predicted outputs, which is particularly well suited to modeling galaxy bias given its highly stochastic nature. The NF model learns a family of conditional probability distributions $p(b_i|\{x\}_i)$, where $i$ indexes individual galaxies and $\{x\}_i$ denotes the set of selected input properties used to predict the individual bias of the galaxy (see Sec. \ref{sec:Data}). This approach enables us to model, for each galaxy in the sample, a full probability distribution of bias values rather than a single point estimate. We refer to these as individual probability distributions, and their shape depends on the selected input features. These distributions—examined in detail in this section—are obtained by sampling 1000 realizations per galaxy from the learned conditional model.

The shape of these individual probability distributions is sensitive to the choice of input properties, varying according to the information provided to the model. To explore this dependence, we compare the impact of two representative inputs: an internal halo property with low importance according to Table~\ref{table:feature importance summary}, being this $\lambda_{\text{halo}}$, and the most influential feature, $\delta_8$. Using the same NF architecture exhibited in Table~\ref{table:hyperparameters NF}, we trained the model three times: (i) using only $\lambda_{\text{halo}}$, (ii) using only $\delta_8$, and (iii) using the full set of selected properties. Figure~\ref{fig:distributions_NF_onegalaxy} shows an example of the resulting individual bias distribution for the same randomly chosen galaxy under these different input configurations in purple, the blue dash line corresponds to the true bias value of the same galaxy, while the cyan distributions correspond to the true individual bias values histograms for the entire test set (as shown in pink in Fig.~\ref{fig:3_models_comparison}).

We find that when using features with low predictive power, such as $\lambda_{\text{halo}}$, and therefore limited information content, the NF tends to reproduce a distribution similar to the overall distribution of the true bias values (see the left panel in Fig.~\ref{fig:distributions_NF_onegalaxy})), meaning that the model is unable to meaningfully constrain the conditional distributions on a per-object basis, particularly in the tails. This behavior arises because the conditioner network is unable to infer much information from this single property and therefore outputs similar spline parameters for all galaxies, resulting in nearly identical conditional distributions ($p(b_i|\lambda_{\text{halo}})$). This effect becomes progressively less pronounced as more informative features are introduced, such as $\delta_8$, and is further mitigated when additional properties are included, as the model can extract more information from the inputs to infer object-dependent spline parameters and better constrain the conditional distributions, leading to variations in the shapes of these distributions that reflect the differences between the galaxies. 

It is noteworthy that, regardless of the chosen input features for the generative model, selecting a single random realization from the full set of samples yields an overall distribution of bias values that closely matches the true one, as shown in the top panel of Fig.~\ref{fig:different_distributions_NF}. 

However, caution must be exercised, as predictions for individual galaxies can be inaccurate when the input features are weakly informative, which is clearly visible when taking the mean value across all sampled realizations. This behavior is illustrated in the bottom panel of Fig.~\ref{fig:different_distributions_NF}, where the contrast between models trained with less informative versus more influential features is evident. In particular, conditioning only on $\lambda_{\text{halo}}$ leads to mean values that cluster around the global average, producing the narrow peak observed in the figure (purple line).

\begin{figure}
    \centering
    \includegraphics[width=0.9\linewidth]{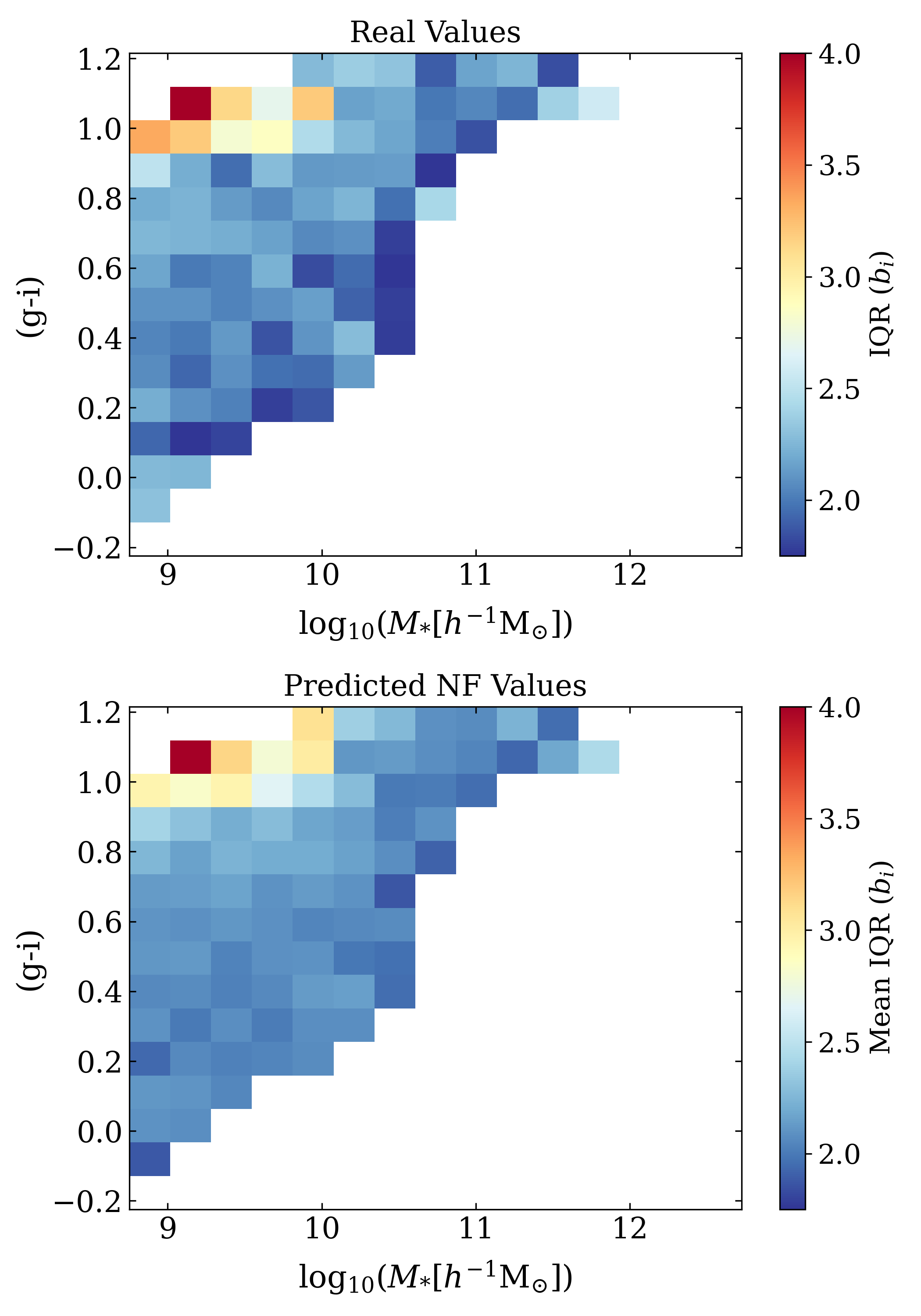}
    \caption{A 2D colormap of the IQR of the bias distributions per pixel. Each plot is divided into a $15\times15$ grid, and only pixels containing 30 objects or more are considered. The top panel shows the IQR of the true bias values for all objects within each pixel, while the bottom panel displays the mean IQR obtained by averaging the IQRs of the individual bias distributions of the galaxies contained in each pixel.}
    \label{fig:plots_pixeles}
\end{figure}

Since NFs can be used to produce bias distributions for individual objects, it is interesting to evaluate how the internal variance of specific galaxy populations can be recovered. To this end, we placed galaxies in the color–stellar mass plane (features that were not predicted in this case) divided into $15 \times 15$ pixels, as shown in Fig.~\ref{fig:plots_pixeles}, where only pixels containing 30 objects or more were considered. For the true bias values, we computed the interquartile range (IQR) of the bias distribution for all objects within each pixel. For the predicted values, we used the full set of properties listed in Sec.~\ref{sec:Data} as input. Since in this case each galaxy in a pixel has its own bias distribution, we calculated the IQR for each individual distribution and then took the mean of all these IQR values. This mean IQR corresponds to the color scale shown for the predicted case in Fig.~\ref{fig:plots_pixeles}. 

\begin{figure}
    \centering
    \includegraphics[width=0.8\linewidth]{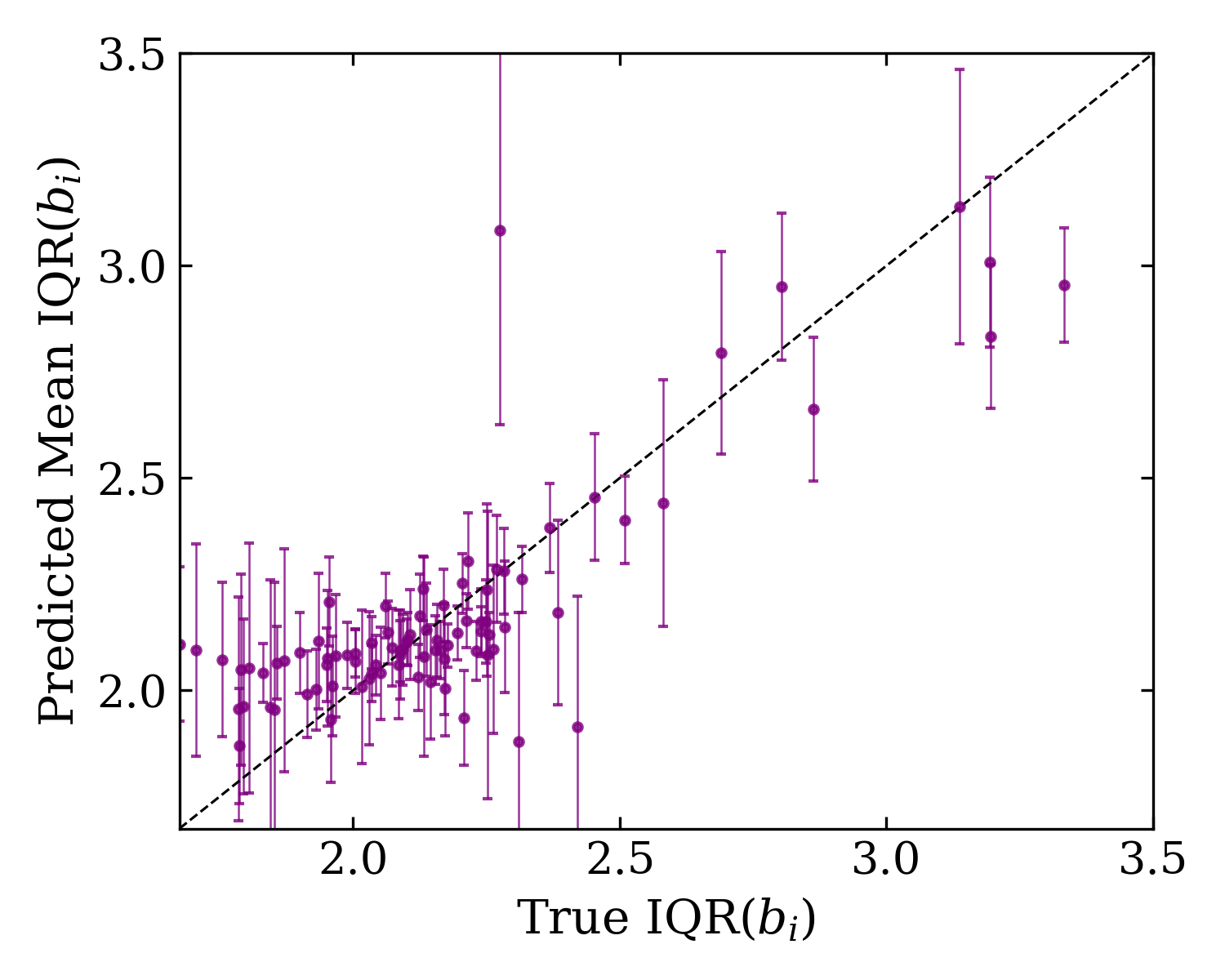}
    \caption{Scatter plot of the predicted mean IQR as a function of the true IQR. Each purple dot represents a pixel (see Fig.~\ref{fig:plots_pixeles}), and the error bars correspond to the standard deviation of the predicted IQR within each pixel. The black dashed line indicates the ideal one-to-one relation ($x = y$), where the predicted and true IQRs are equal.}
    \label{fig:scatterplot_pixeles}
\end{figure}

The first notable result is the clear trend of the IQR increasing diagonally toward the red, low-mass corner of the parameter space, in both the real and the predicted case, which exhibit similar range of IQR values according to Fig.~\ref{fig:plots_pixeles}. To visualize this comparison more clearly, Fig.~\ref{fig:scatterplot_pixeles} displays a scatter plot of the predicted mean IQR as a function of the true IQR, along with the corresponding error bars, representing the standard deviation of the predicted IQR within each pixel. Although some discrepancies between the two cases exist, they follow the same general trend: the larger the dispersion in the data, the larger the predicted one. This result demonstrates that NFs can effectively capture the scatter of the bias parameter when using halo and environmental features as input, representing an important step toward characterizing its stochastic nature.

\section{Summary and conclusions}
\label{sec:Discussion and Conclusions}
In this work, we present a ML framework aimed at reproducing the bias of galaxies as a function of internal halo properties and environmental diagnostics, as well as information on the location of galaxies within the cosmic web provided by DisPerSE. One of the main innovations of this framework is the adoption of the prescription of \citet{Paranjape2018} and \cite{Balaguera2024A} to assign an individual bias value to each galaxy in the TNG300 hydrodynamical simulation, allowing us to incorporate bias into the ML machinery as an additional object-by-object property. We subsequently apply three different ML algorithms: two deterministic single-point estimators (a Random Forest Regressor and a single-output Neural Network) and one probabilistic approach (Normalizing Flows) to predict the individual bias values following the methodologies of \citet{de_Santi_2022} and \citet{Rodrigues_2023, Rodrigues2025}. The performance of all models is compared using three complementary evaluation metrics: the Pearson Correlation Coefficient, the Kolmogorov–Smirnov test, and the Wasserstein Distance. Finally, we study the individual conditional probability distributions provided by the NF and analyze the ability of this generative method to reproduce the intrinsic variance of the bias parameter. The main conclusions of this paper can be summarized as follows: 

\begin{itemize}
 
\item Using a correlation analysis, we show that galaxy bias exhibits strong correlations with the overdensities measured on different scales ($\delta_3$, $\delta_5$, and $\delta_8$). These correlations strengthen with increasing scale, with $\delta_8$ emerging as the most strongly correlated feature. For sufficiently large $R$, both $\delta_R$ and $b_i$ probe a similar underlying quantity \citep{Paranjape2018}, which explains this tight connection (see also \citealt{Pujol2017, 2023A&A...673A.130B, MonteroDorta2024}). In order of decreasing strength, these correlations are followed by distances to cosmic-web critical points and, finally, by internal halo properties, with $z_{1/2}$ being the most strongly correlated among them.

\item When analyzing the importance of each selected property in predicting galaxy bias across the different ML models, we find that all three models also identify $\delta_8$ as the most important feature, followed by $\delta_5$. Overall, the models are consistent with our correlation analysis in indicating that the environment has a greater impact on predicting the individual galaxy bias than the internal halo properties. The most notable difference is that $\delta_3$ becomes mostly uninformative once the larger-scale overdensities are included in the analysis.

\item We have shown that deterministic estimators are inherently limited in their ability to reproduce the true galaxy bias distributions with high fidelity, given the stochastic nature of the halo-galaxy connection. \citet{Rodrigues2025} proposes generative models to address this problem, and our results also demonstrate that NFs provide an effective approach to predict galaxy bias. In the bias-only case, the model achieves a K-S statistic of 0.010 and a Wasserstein Distance of 0.046, considering a random realization. When predicting joint probability distributions, the model reaches a 2-D K-S statistic of 0.017 and a 2-D WD of 0.030 for the $b_i$–$\log_{10}(M_*[h^{-1}\text{M}_{\odot}])$ relation, and a 2-D K-S statistic of 0.029 and a 2-D WD of 0.052 for the $b_i$–$(g-i)$. 

\item We find that when conditioning the NF model on properties that exhibit little or no correlation with the individual galaxy bias, such as halo spin, the model struggles to properly constrain the conditional individual probability distributions. In this case, the NF lacks sufficient information to distinguish between galaxies and therefore the conditioner network assigns similar conditional distributions to all objects. In contrast, the predicted distributions change significantly as more informative inputs are included, particularly features that are strongly correlated with the bias and therefore have greater predictive power.

\item We demonstrate that the NF predictions capture the behavior of the dispersion of the real data in a color--mass diagram, by following the same trend: the larger the variance in the true values, the larger the predicted one. These results, combined with the 1D and 2D distributions, demonstrate that such techniques are well suited to recover the intrinsic scatter of galaxy bias, reflecting the underlying complexity of the physical processes driving halo and galaxy formation and evolution.

\end{itemize}

Although the performance of all employed methods reflects the inherent challenges of predicting stochastic parameters, the predictive power of each model in this context also depends on the selected features. As future work, it would be interesting to evaluate the models’ performance when incorporating additional properties, for example, by incorporating more complex evolutionary information using the merger trees (e.g., \citealt{Jespersen2022}) and employ Graph Neural Networks (GNNs; \citealt{Zhou_2021}) in the context of bias prediction. Second, we could test our results against additional sophisticated ML models, as well as testing these models in larger simulations, for example the MilleniumTNG simulation \citep{Hernandez-Aguayo_2023}. Finally, it is possible to incorporate additional environmental diagnostics, like a refined cosmic web classification scheme \citep{Galarraga_Espinosa_2023} or the inclusion of sophisticated anisotropy estimators (Riveros-Jara, in prep.).

This work is expected to have important future applications. First, evaluating the ability of probabilistic approaches to capture the variance (i.e., intrinsic scatter) of galaxy bias based on different halo and environmental properties represents an important step toward assessing the potential of this framework to fully characterize bias stochasticity, including through the exploration of additional ML models and different simulations. Second, these results constitute a meaningful advancement to measure galaxy bias observationally using data from upcoming surveys, such as the Dark Energy Spectroscopic Instrument (DESI; \citealt{DESI2016}). In this context, the use of mock catalogs to test the applied models becomes essential before using these techniques on real data.

\begin{acknowledgements}
CRJ aknowledges Ricardo Ñanculef and the Department of Informatics of the Universidad Técnica Federico Santa María (UTFSM) for the computational resources provided for this project. CRJ and ADMD acknowledge support from the UTFSM through the Proyecto Interno Regular \texttt{PI\_LIR\_25\_04}. ABA acknowledges the Servicio Público de Empleo del Gobierno de España. NSMS acknowledges partial support from the NSF CDSE grant AST-2408026 and the NASA TCAN grant 80NSSC24K0101. MCA acknowledges partial support from ANID BASAL project FB210003.
\end{acknowledgements}

\bibliographystyle{aa} 
\bibliography{references}
\begin{appendix}
\twocolumn
\section{Coverage Test}
\label{appendix:Apendix_2}
In order to assess the accuracy of the NF predictions, we performed a coverage test utilizing the TARP test. The coverage test evaluates the accuracy and calibration of probabilistic predictions by verifying whether the estimated posterior distributions correctly contain the true values at the expected credibility levels.

For each data point, many samples are drawn from the predicted posterior and compared to the true value from the simulation. The test measures the expected coverage probability (ECP) of credible regions at different credibility levels. If the posterior estimator is accurate, the empirical coverage should match the nominal credibility, yielding a one-to-one relation between coverage and credibility. Deviations from this diagonal indicate miscalibration: under-coverage signals overconfident or biased posteriors, while over-coverage indicates underconfident or overly broad predictions. A more detailed description of this test can be found in \cite{Rodrigues2025}.

\begin{figure}[h!]
    \centering
    \includegraphics[width=0.8\linewidth]{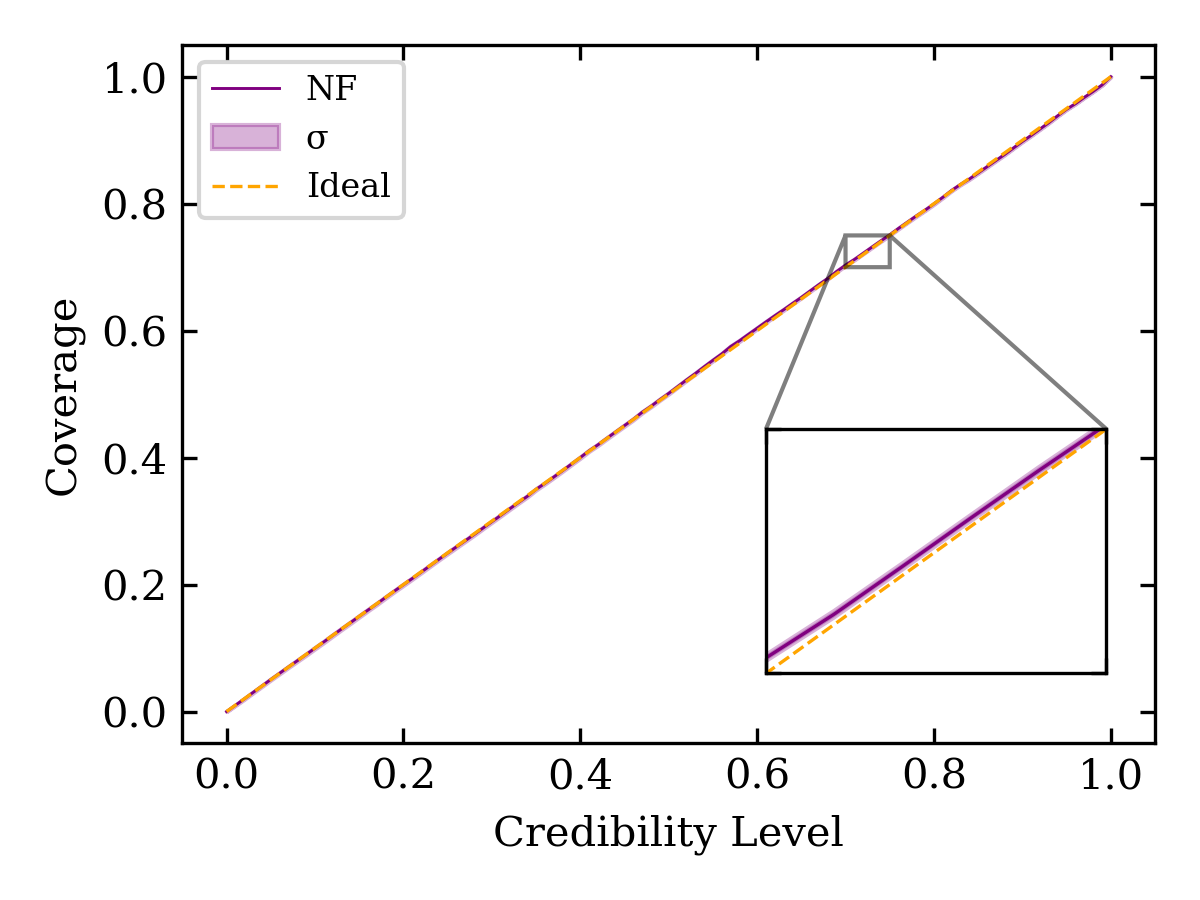}
    \caption{TARP coverage test for the 1-D case predictions with NF. The dashed orange line corresponds to the ideal one-to-one relation. The purple line exhibits the mean TARP for the model while the purple area covers one $\sigma$ around the mean value. }
    \label{fig:tarp test}
\end{figure}

Figure~\ref{fig:tarp test} shows the coverage test for the bias-only (1-D case) predictions made with the NF model on the test set, suggesting our model is well calibrated, as it only presents small deviations from the one-to-one expected correspondence. 

\section{Correlation between features}
\label{appendix:Apendix_1}
\begin{figure*}[h!]
    \centering
    \includegraphics[width=0.9\textwidth]{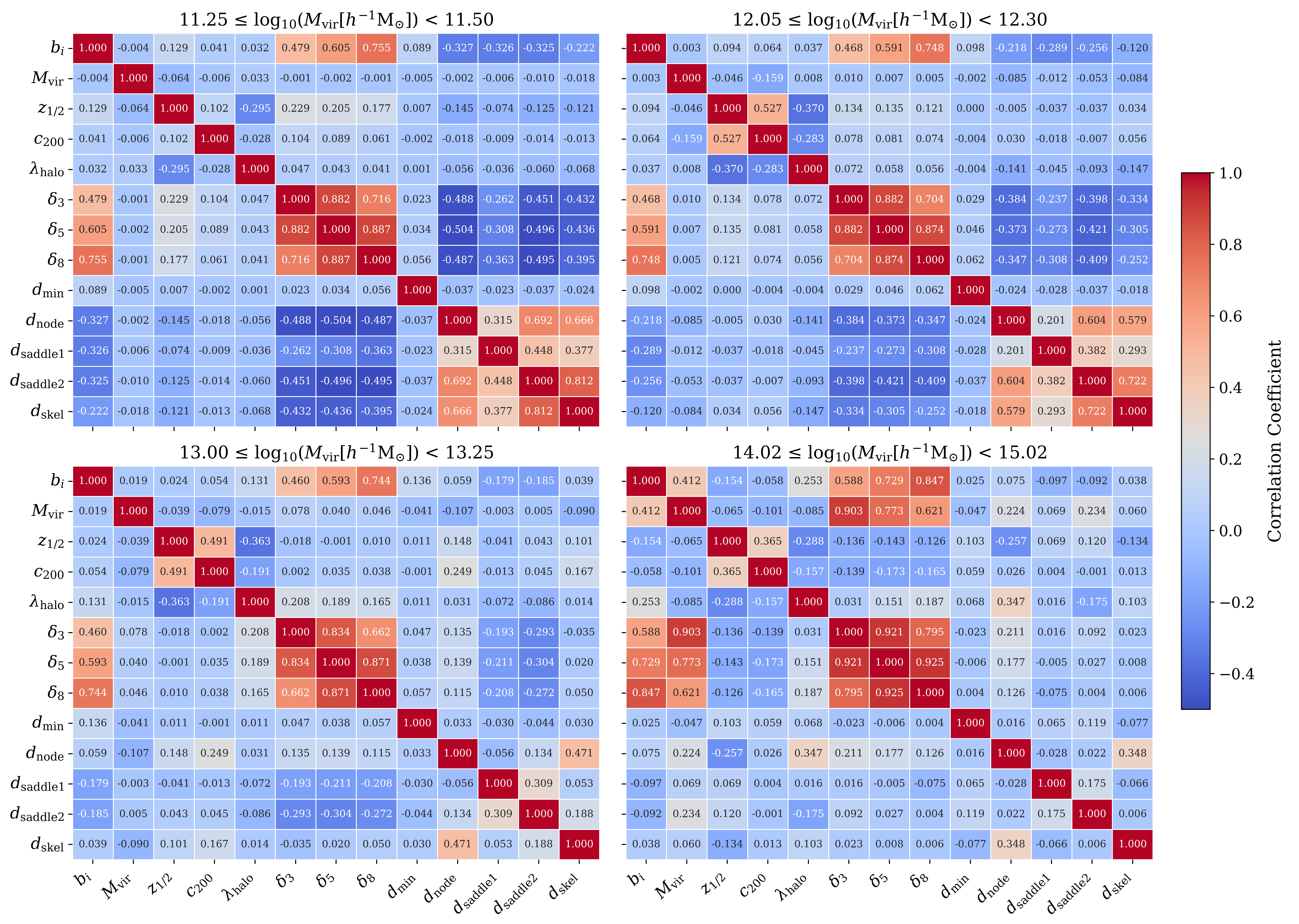}
    \caption{Complete correlation matrices for the four different mass bins presented in Fig.~\ref{fig:corr_matrix_bins_puntos}. Each plot shows the linear correlation between all features within the given mass bin.}
    \label{fig:corr_matrix_bins}
\end{figure*}
One of the main advantages of adopting an object-by-object prescription to estimate the large-scale linear bias for each galaxy in a given sample (in this work, within the TNG300 simulation box) is that it allows us to disentangle its dependence on various properties, such as those related to the host halos and the local environment. In this work, we begin by analyzing how each property in the catalog correlates with the individual galaxy bias ($b_i$), as shown in Fig.~\ref{fig:corr_matrix_resumen} and Fig.~\ref{fig:corr_matrix_bins_puntos} for different halo mass bins. However, there are additional interesting relations among the selected properties that may be of interest to the reader and/or could be explored in future studies using the same techniques proposed here. For this reason, to complement the analysis presented in Sec.~\ref{sec:Correlation Between Features}, here we describe the full correlation matrices in Fig.~\ref{fig:corr_matrix_bins}, which show the linear correlations within the four distinct halo mass bins described in Sec.~\ref{sec:Correlation Between Features}. In this appendix, we focus on the relationships among the features excluding $b_i$, as the correlations involving the bias are already discussed.

A notable trend in Fig.~\ref{fig:corr_matrix_bins} is the evolution of the correlation between the overdensities ($\delta_3$, $\delta_5$, and $\delta_8$) and the distances to critical points ($d_{\text{node}}$, $d_{\text{saddle1}}$, $d_{\text{saddle2}}$, and $d_{\text{skel}}$). For the lower mass bins, these correlations are predominantly negative, whereas they shift toward positive values as the halo mass increases. This behavior suggests a mass-dependent correlation between the overdensities and the distances to critical points, which becomes more pronounced for more massive halos. In contrast, no clear trend is observed in the correlation between the overdensities and the distance to the nearest void ($d_{\text{min}}$).

Small variations are found in the correlations involving $z_{1/2}$ and $c_{200}$, which increase up to the second mass bin and then decrease for higher masses. A similar behavior is observed for their anti-correlation with the halo spin parameter.

\section{Multiproperty predictions}
\label{sec:multiproperty predictions}
\begin{figure*}
    \centering
    \includegraphics[width=0.8\linewidth]{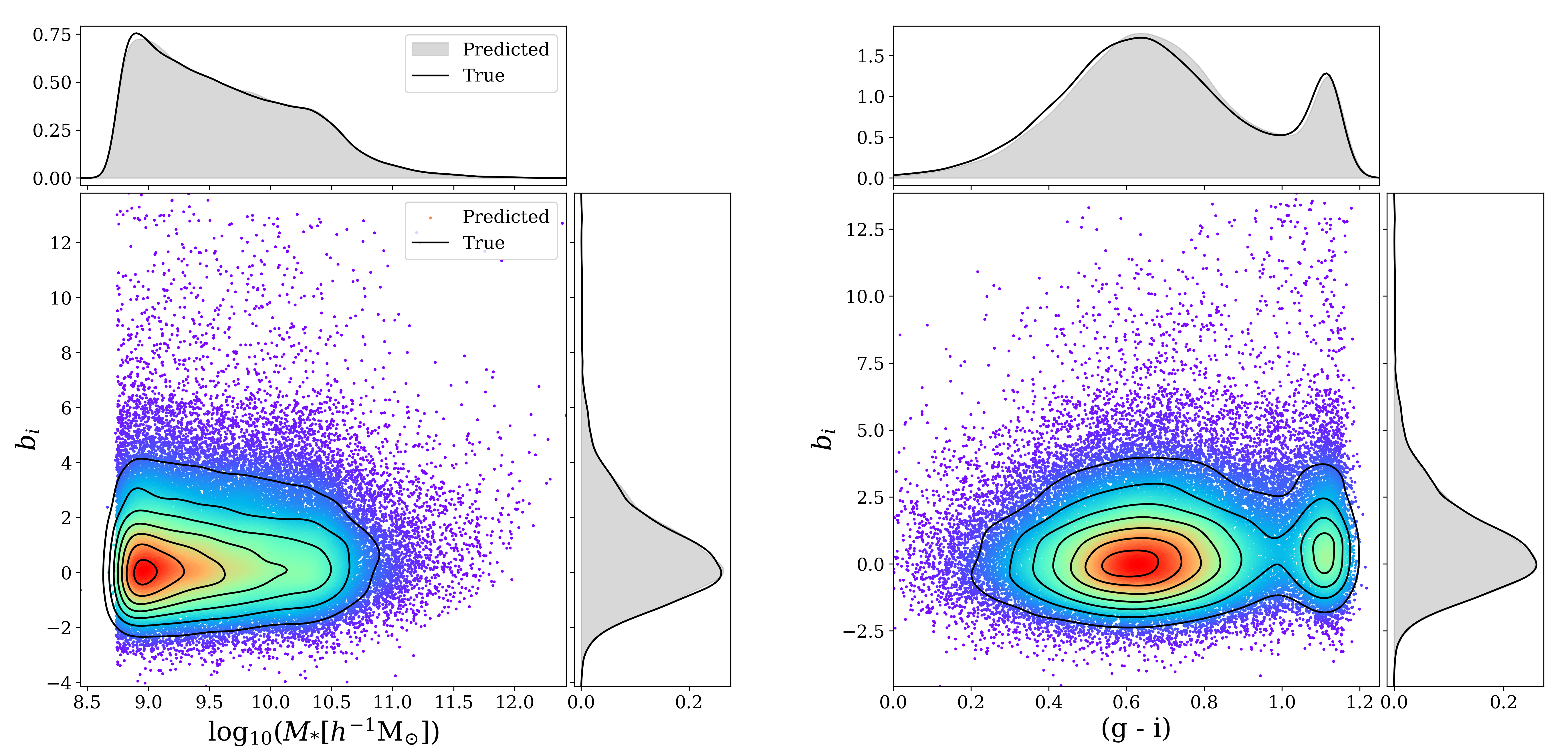}
    \caption{Joint probability distribution of both individual galaxy bias and stellar mass (left panel), as well as individual galaxy bias and galaxy color (right panel). The main panels show predicted values from one random realization, while black contour lines represent the density of the true values. The panels on the top and to the right of the main ones show the marginal density distributions for predicted values (gray) and true values (black lines).}
    \label{fig:joint_dist}
\end{figure*}
 Following the approach of \cite{Rodrigues_2023,Rodrigues2025}, we jointly predict two pairs of properties as relevant examples: (i) the individual galaxy bias and stellar mass ($p(\{b_i, M_*\}|\{x\}_i$), and (ii) the individual galaxy bias and galaxy color ($p(\{b_i, g-i\}|\{x\}_i$), adopting the same architecture presented in Table~\ref{table:hyperparameters NF}. These choices are motivated by two main considerations. First, both properties are observationally measurable. Second, the stellar mass of a galaxy is strongly correlated with the mass of its host halo, and galaxy color is linked to its accretion history \citep{Hearin2013,Wechsler2018, Behroozi2019, ChavesMontero2020, MonteroDorta2021}. 

The predictions for both cases are presented in Fig.~\ref{fig:joint_dist}, obtained by selecting a single random realization from the sampled catalog. Here, the main panels display the predicted values, color-coded by normalized density, while the contour lines represent the density of the true values from the catalog. The panels located above and to the right of each main panel show the marginal density distributions for both the predicted (gray area) and true (black lines) values. All densities (both in the scatter plots and histograms) were estimated using Gaussian KDE from the \texttt{scipy.stats} module.

To quantify the difference between the predicted and true values for these multiproperty cases, we used the 2-D K-S Test and the 2-D Wasserstein Distance. The obtained values of each metric on the test set are listed in Table~\ref{table:multiproperty metrics}. 
\begin{table}[h!]
\caption{Metric values for the multiproperty case obtained with the NF model.}                 
\label{table:Metrics joint-prob}    
\centering                     
\begin{tabular}{c c c}      
\hline\hline               
Predicted with $b_i$ & 2-D K-S Test & 2-D WD\\       
\hline                 
   $\log_{10} (M_*[h^{-1} \text{M}_{\odot}])$ & 0.017 & 0.030\\   
   $(g-i)$& 0.029 & 0.052\\
\hline   
\label{table:multiproperty metrics}
\end{tabular}
\end{table}

The low values listed in Table~\ref{table:multiproperty metrics} demonstrate that NFs can effectively and accurately predict joint features, proving to be a powerful framework capable of predicting multiple correlated properties jointly rather than a single one. Although this study focuses on pairs of features, the method can be extended to include higher-dimensional joint predictions\footnote{In this context, it is straightforward to  extend our framework to predict secondary bias.} (see further discussion in \citealt{Rodrigues2025}).

A key result of this analysis is that NFs successfully reproduce the scatter in the bias–stellar mass and bias–color relations presented in Fig. \ref{fig:joint_dist}. This scatter at fixed stellar mass and color reflects the underlying complexity of the physical processes driving halo and galaxy formation and evolution.

\end{appendix}

\end{document}